\documentclass[sigconf]{acmart}
\AtBeginDocument{%
  \providecommand\BibTeX{{%
    \normalfont B\kern-0.5em{\scshape i\kern-0.25em b}\kern-0.8em\TeX}}}
\copyrightyear{2023} 
\acmYear{2023} 
\setcopyright{acmlicensed}\acmConference[WebSci '23]{15th ACM Web Science Conference 2023}{April 30-May 1, 2023}{Evanston, TX, USA}
\acmBooktitle{15th ACM Web Science Conference 2023 (WebSci '23), April 30-May 1, 2023, Evanston, TX, USA}
\acmPrice{15.00}
\acmDOI{10.1145/3578503.3583619}
\acmISBN{979-8-4007-0089-7/23/04}

\usepackage{graphicx}
\usepackage{enumitem}
\usepackage{xspace}
\usepackage{subcaption}
\usepackage{amsmath}
\usepackage{array}
\usepackage{multicol}
\usepackage{multirow}
\newcommand{\squishlist}{
 \begin{list}{$\bullet$}
  { \setlength{\itemsep}{1pt}
     \setlength{\parsep}{0pt}
     \setlength{\topsep}{0pt}
     \setlength{\partopsep}{20pt}
     \setlength{\leftmargin}{1.5em}
     \setlength{\labelwidth}{1em}
     \setlength{\labelsep}{0.5em}}}
\newcommand{\squishend}{
  \end{list}  }

\title{A longitudinal study of the top 1\% toxic Twitter profiles}

\author{Hina Qayyum}
\affiliation{%
  \institution{Macquarie University}
  \city{Sydney}
  \country{Australia}}
\email{hina.qayyum@mq.edu.au}

\author{Benjamin Zi Hao Zhao}
\affiliation{%
  \institution{Macquarie University}
  \city{Sydney}
  \country{Australia}}
\email{ben\_zi.zhao@mq.edu.au}

\author{Ian D. Wood}
\affiliation{%
  \institution{Macquarie University}
  \city{Sydney}
  \country{Australia}}
\email{ian.wood@mq.edu.au}

\author{Muhammad Ikram}
\affiliation{%
  \institution{Macquarie University}
  \city{Sydney}
  \country{Australia}}
\email{muhammad.ikram@mq.edu.au}

\author{Mohamed Ali Kaafar}
\affiliation{%
  \institution{Macquarie University}
  \city{Sydney}
  \country{Australia}}
\email{dali.kaafar@mq.edu.au}

\author{Nicolas Kourtellis}
\affiliation{%
  \institution{Telefonica Research}
  \city{Barcelona}
  \country{Spain}}
\email{nicolas.kourtellis@telefonica.com}



\begin{abstract}
Toxicity is endemic to online social networks (OSNs) including Twitter. It follows a Pareto-like distribution where most of the toxicity is generated by a very small number of profiles and as such, analyzing and characterizing these ``toxic profiles'' is critical. Prior research has largely focused on sporadic, event-centric toxic content (i.e., tweets) to characterize toxicity on the platform. Instead, we approach the problem of characterizing toxic content from a profile-centric point of view.
We study 143K Twitter profiles and focus on the behavior of the top 1\% producers of toxic content on Twitter, based on toxicity scores of their tweets availed by Perspective API. With a total of 293M tweets, spanning 16 years of activity, the longitudinal data allows us to reconstruct the timelines of all profiles involved.
We use these timelines to gauge the behavior of the most toxic Twitter profiles compared to the rest of the Twitter population. 
We study the pattern of tweet posting from highly toxic accounts, based on the frequency and how prolific they are, the nature of hashtags and URLs, profile metadata, and Botometer scores. We find that the highly toxic profiles post coherent and well-articulated content, their tweets keep to a narrow theme with lower diversity in hashtags, URLs, and domains, they are thematically similar to each other, and have a high likelihood of bot-like behavior, likely to have progenitors with intentions to influence, based on high fake followers score.
Our work contributes insight into the top 1\% toxic profiles on Twitter and establishes the profile-centric approach to investigate toxicity on Twitter to be beneficial. The identification of the most toxic profiles can aid in the reporting and suspension of such profiles, making Twitter a better place for discussions.  Finally, we contribute to the research community with this large-scale and longitudinal dataset\footnote{https://github.com/hqayyum/twitter\_top\_toxic\_1percent}, annotated with six types of toxic scores.
\end{abstract}

\begin{document}
\maketitle
\section{Introduction}
\label{sec:intro}
Verbal misbehavior and toxicity on Online Social Networks (OSNs) are receiving a huge amount of attention in the research community, with efforts to identify~\cite{zhang2019hate, Gaydhani2018DetectingHS, 8292838, basile-etal-2019-semeval, calderon2020topic}, characterize~\cite{Criss202, clarke2017dimensions, benigni2017online, fernandez2018understanding, chatzakou2017mean}, and automatically detect  ~\cite{anzovino2018automatic, fortuna2018survey, d2020bert} online misbehavior, especially on Twitter. 
Despite all these ongoing efforts, toxicity has increased over time. We note that almost all efforts to study toxicity on Twitter come from the content study of tweets posted around sporadic high-profile campaigns and events such as elections~\cite{https://doi.org/10.48550/arxiv.2103.01664}, important world events of COVID19 and the MeTooMovement~\cite{https://doi.org/10.48550/arxiv.2005.12423,ALkhalifa2020}, or controversies about topics like Bitcoin~\cite{pacheco2021uncovering}. 
However, these studies do not explore the prolonged involvement of a profile in spreading toxic content, so its utility in the characterization of overall toxicity was hindered. 
Works like~\cite{Wei2018,https://doi.org/10.48550/arxiv.1803.08977} investigate toxic profiles responsible for disseminating toxic content on small manually annotated datasets. 
In essence, efforts toward the automatic detection and characterization of toxicity on Twitter are mostly event-centric or small-scale, user-centric. This scenario leaves a gap in understanding the entire landscape of misbehavior on Twitter. 

Toxic content follows a Pareto-like distribution on Twitter~\cite{ribeiro2018characterizing}, hence we focus on the most toxic profiles in our dataset based on the median Perspective ``Toxicity'' score of the profile's tweets. We contrast these profiles with the remainder of our dataset to find out how much their behavior is different from base Twitter profiles. We focus on research questions that will assist us in better understanding toxic profiles:
\begin{itemize}[leftmargin=*]
    \item Are toxic profiles prolific content generators, with a specific tweeting pattern?
    \item Do toxic profiles tweet in a legible way to effectively convey their message?
    \item What type of misconduct is expected of a toxic profile?
    \item Do toxic profiles leverage auxiliary content, such as URLs and hashtags?
    \item Do toxic profiles demonstrate special trends with respect to name, location, counts of friends or followers, and such?
    \item Can we expect very toxic profiles to be bad bots?
\end{itemize}
Our dataset, described in \S\ref{sec:dataset}, is seeded with seven smaller public datasets from past works studying online misbehavior on Twitter. These seed datasets cover  
multiple themes of online misbehavior: hostility, racism, abuse, hatefulness, homophobia, spam, and sexism, and are balanced in the number of toxic and non-toxic users. A key limitation of the seed datasets is that users are classified as toxic or not from the content of a single or a few tweets, which does not allow deeper analysis of the users' average toxic behavior. To enable such analysis, we crawl the tweet timeline of each of the users present in the seed datasets. 
Our resulting dataset contains 142,987 (143K) Twitter profiles and 293,401,161 (293M) individual tweets posted between 2007 and 2021. Human annotations are 
untenable given the size of our dataset,
hence, we turn to Google's Perspective APS~\cite{perspective} models to assign toxicity scores to each tweet, providing estimates of the following types of misbehavior: 
\emph{Toxicity, Severe Toxicity, Identity Attack, Inflammatory, Threat, and Insult}. We use only the production-ready scores from the Perspective API, which provide highly reliable estimates. 

In \S\ref{sec:prolificacy_analysis}, we investigate these highly toxic profiles with respect to tweeting frequency and dynamics, drawing on the distribution of inter-tweet times and a measure of burstiness.
Next in \S\ref{sec:content_analysis}, we look at the tweet content. We explore the Perspective scores  
and their consistency among each profile's tweets using the Gini Index~\cite{gini1912variabilita}. Next, we study the number and quality of hashtags and URLs with help of Fortiguard, a service that categorizes URLs by topic. We then perform a readability analysis of tweets, using Flesch reading ease and difficulty scores, Linsear write score, and the Automatic Readability Index (ARI). In \S\ref{sec:profile_analysis}, we characterize toxic profiles based on the profile metadata including the number of friends, followers, statuses, favorites, membership of lists, location, creation date, and profile status. In addition, we use to apply the Botometer API~\cite{Sayyadiharikandeh_2020} to our profiles, obtaining scores that quantify astroturfing, spamming, fake followers, self-declared bots, and financial bots. 

This work makes the following main contributions: 
\begin{itemize}[leftmargin=*]
    \item We collect and curate a longitudinal dataset of tweets, spanning 16 years, consisting of 293M tweets (\S\ref{sec:dataset}) and augmented with six perspective scores. To our knowledge, this is the largest annotated dataset on online misbehavior. To foster further research, upon publication, we plan to share our dataset with the research community.
    \item We identify that the top 1\% toxic profiles post fewer, shorter, but more articulate tweets than the rest. 
    We find that the Gini Index of Perspective scores on each toxic profile's tweets is lower, indicating consistency of misbehavior among their tweets.  
    \item We observe that the highest Perspective scores among 
    tweets from toxic profiles are ``Inflammatory'' and ``Insult'', and that ``Identity attack'' is relatively low, especially when compared to where it sits among baseline tweets. 
    \item The top 1\% toxic profiles tend to use fewer but more coherent and similar URLs and hashtags.
    \item We find that the top 1\% toxic profiles have lower friends and follower counts than baseline profiles.
    \item We observe a notable increase in the creation of toxic profiles between 2014-2016. Interestingly, we note that despite being the most toxic, none of the top 1\% profiles has been deactivated, banned, or deleted in the 18 months that passed between timeline data crawling and profile metadata extraction. 
    \item Notably, we identify that just under half the top 1\% toxic profiles would be classified by Botometer as fake followers, which is important evidence of instrumented trolling campaigns on Twitter.
\end{itemize}

\section{Related Work}
\label{sec:rwork}
Online misbehavior detection on social networks and on Twitter has been extensively explored by several studies such as~\cite{gomez2019exploring,ribeiro2018like,founta2018large, waseem-hovy-2016-hateful, DHUNGANASAINJU2021106735}. However, almost all approaches to indicate hate or toxicity is content-centric, the inherent shortfall of collecting and annotating toxic tweets is due to the incompleteness and insufficiency of OSN text i.e., tweets, and the sparsity of toxic hateful speech. These limitations are often amplified by oversimplifying the problem, such as considering only tweets collected around extremist events or collected with keyword searches. In this work, we partially address these limitations by accumulating a user-centric dataset, such work is done on a very limited level by ~\cite{https://doi.org/10.48550/arxiv.1803.08977}. 

Past studies \cite{gomez2019exploring, ribeiro2018like} have relied on human annotations to differentiate between toxic and non-toxic tweets. Our work leverages the ML models of the Perspective API to rate the collected tweets to explore the different dimensions of misbehavior beyond the prior works, and at the larger scale of data, i.e., 293M tweets. Similarly, previous work \cite{hosseini2017deceiving, jain2018adversarial} has studied Google's Perspective API~\cite{perspective} and its resilience against adversarial attacks. Those studies leveraged Perspective API to score and analyze the toxicity of tweets. 
This work takes precedence over these studies in terms of the size of the data set (293M tweets) and the number of misbehavior dimensions not studied in the past, namely Insult, Inflammatory, Threat, and Identity Attacks.

Automated accounts, paid bots, or trolls' role in toxic and false content creation and dissemination~\cite{PewResearchCenter2018, Ferrara2020, TwitterBotAccounts},  is the base of a consistent spread of toxicity on OSNs. Content-based features best predict coordinated efforts of these malefactors ~\cite{FFI-RAPPORT}, but unsupervised ML for detection of coordinated efforts of profiles in carrying these operations are infeasible at scale~\cite{alizadeh2020content}.
Analysis of unlabeled profiles' longitudinal and unlabeled content provides a characterization of the most toxic users and their content on Twitter.

\begin{figure}[!t]
    \centering
    \includegraphics[width=1\linewidth]{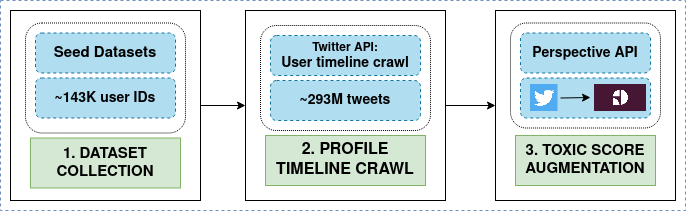}
    \vspace{-4mm}
    \caption{\small Our dataset collection and augmentation pipeline; including dataset collection, user timeline crawl, and augmentation.}
    \label{fig:data_collection_flowchart}
    \vspace{-8mm}
\end{figure}

\section{Dataset Collection Methodology and Characterization}
\label{sec:dataset}
In this section, we detail our methodology for data collection and augmentation. 
We introduce our seed datasets, detail how the timelines of Twitter profiles were crawled, and how we augmented the collected tweet data with Google's Perspective API. We finish with an overall characterization of the augmented dataset, and our definition of the top 1\% toxic Twitter profiles i.e., the upper echelon of toxic profiles as determined by the ``Toxicity'' Perspective score of their tweets.
A summary of the selected datasets, details of their size, and labels can be found in Tab.~\ref{tab:dataset_info} and Tab. \ref{tab:dataset_summary}. Further, a flowchart of our data collection and augmentation methodology is overviewed in Fig.~\ref{fig:data_collection_flowchart}.

\begin{table}[t]
\centering
\resizebox{1.0\linewidth}{!}{
\begin{tabular}{c|r|r|r|p{3.6cm}|p{2cm}}
\hline
\multicolumn{6}{c}{\bf Seed Dataset} \\\midrule

\bf{Dataset} & \bf{TIDs} &  \bf{UIDs} & \bf{RUIDs} & \bf{Labels/Keywords} & \bf{Annotation} \\ \midrule

~\cite{gomez2019exploring} & 149,823 & - & 895 & sexist, racist, homophobic, religion, other hate, no hate    &   Amazon Mechanical Turk    \\ \hline
~\cite{kaggle:metoomovement} & 817,344 & - & 19,859 &  Keyword: MeTooMovement    &  Twitter Streaming API   \\ \hline  
~\cite{ribeiro2018like} & - & 100,385 & 100,385 & hateful, not hateful    &  CrowdFlower (appen) \\ \hline
~\cite{founta2018large} & -  & 98,378 & 98,378 & normal, abusive, spam, hateful    & CrowdFlower (appen) \\ \hline 
~\cite{jha-mamidi-2017-compliment}   & 10,583 & - & 324 & benevolent, hostile, other    & SVM (TF-IDF)    \\ \hline
~\cite{waseem-hovy-2016-hateful} & 16,907 & - & 891 & sexist, racist, neither    &  CrowdFlower (appen) \\ \hline
~\cite{waseem-2016-racist}  & 6,909 & - & 870 & sexist, racist, both, neither    &  CrowdFlower (appen) \\
\bottomrule
\end{tabular}
}
\caption{\small Overview of 7 datasets used as a seed with a collection of User IDs (UIDs) or Tweet IDs (TIDs), whatever was made publicly available.
Note that TIDs were used to recover User IDs (RUIDs).}

\label{tab:dataset_info}
\end{table}
\begin{figure*}[t]
    \centering
    \begin{subfigure}[t]{0.32\linewidth}
            \includegraphics[width=\textwidth]{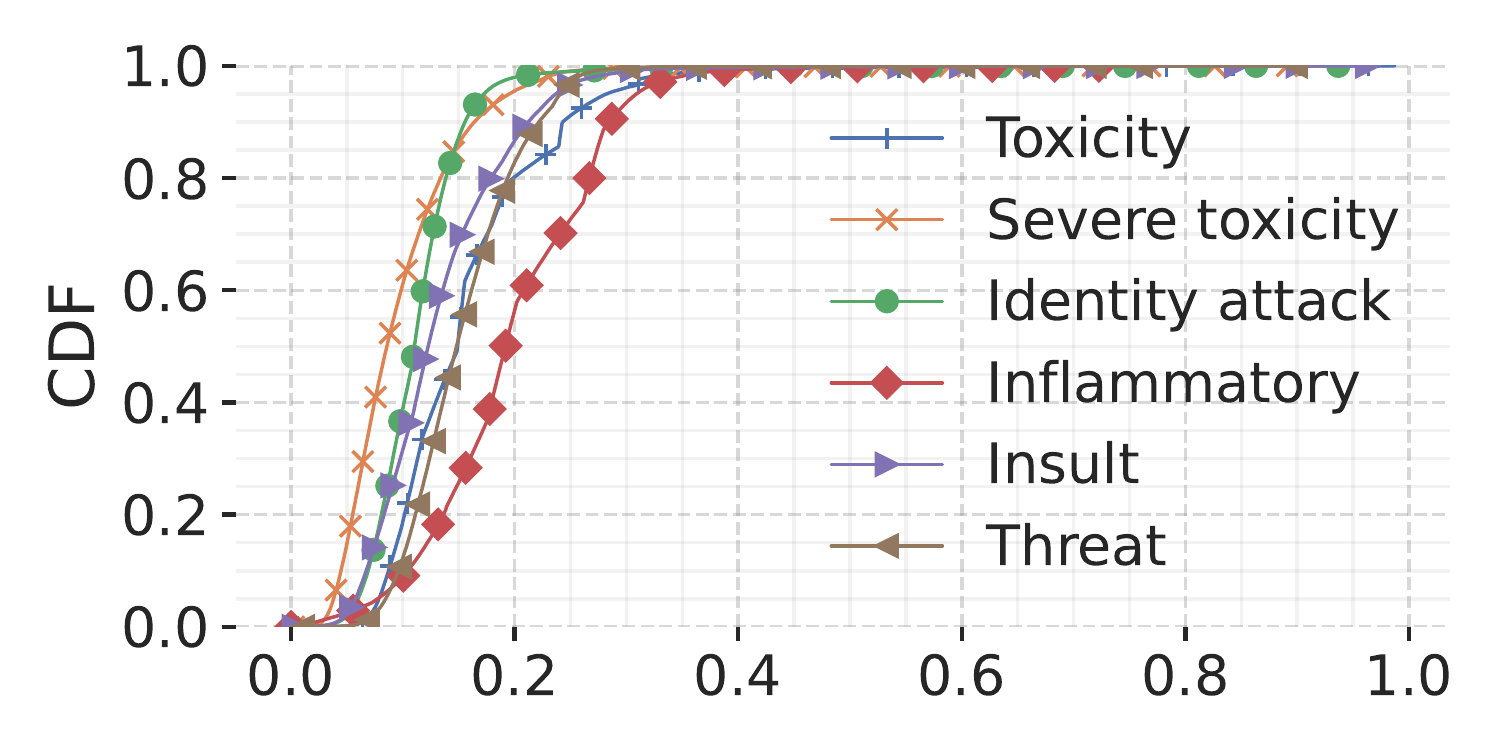}
            \vspace{-6mm}
            \caption{\small Perspective API scores}
            \label{fig:perspective_scores_cdf}
    \end{subfigure}
    \begin{subfigure}[t]{0.32\linewidth}
            \includegraphics[width=\textwidth]{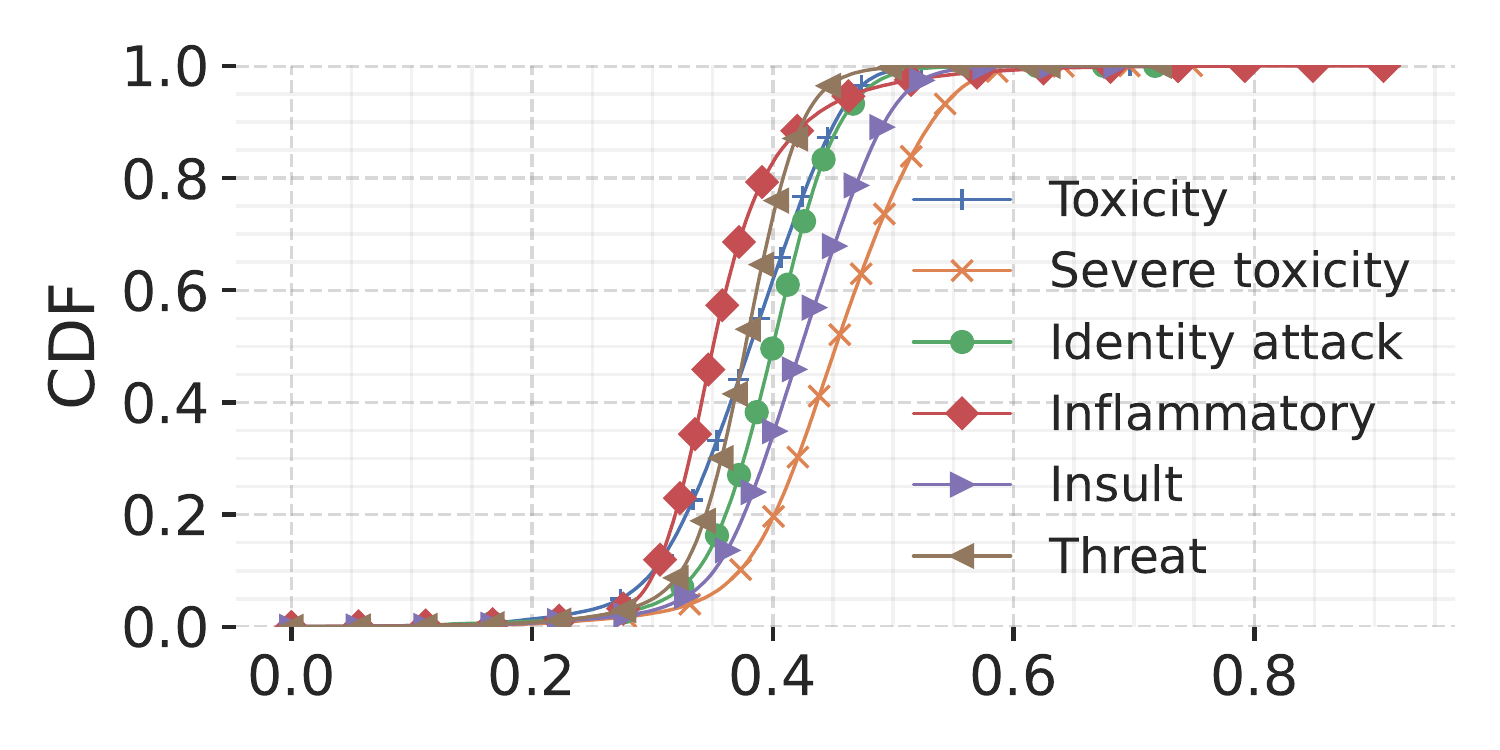}
            \vspace{-6mm}
            \caption{\small Gini index}
            \label{fig:gini_index_cdf}
    \end{subfigure}
    \begin{subfigure}[t]{0.32\linewidth}
            \includegraphics[width=\textwidth]{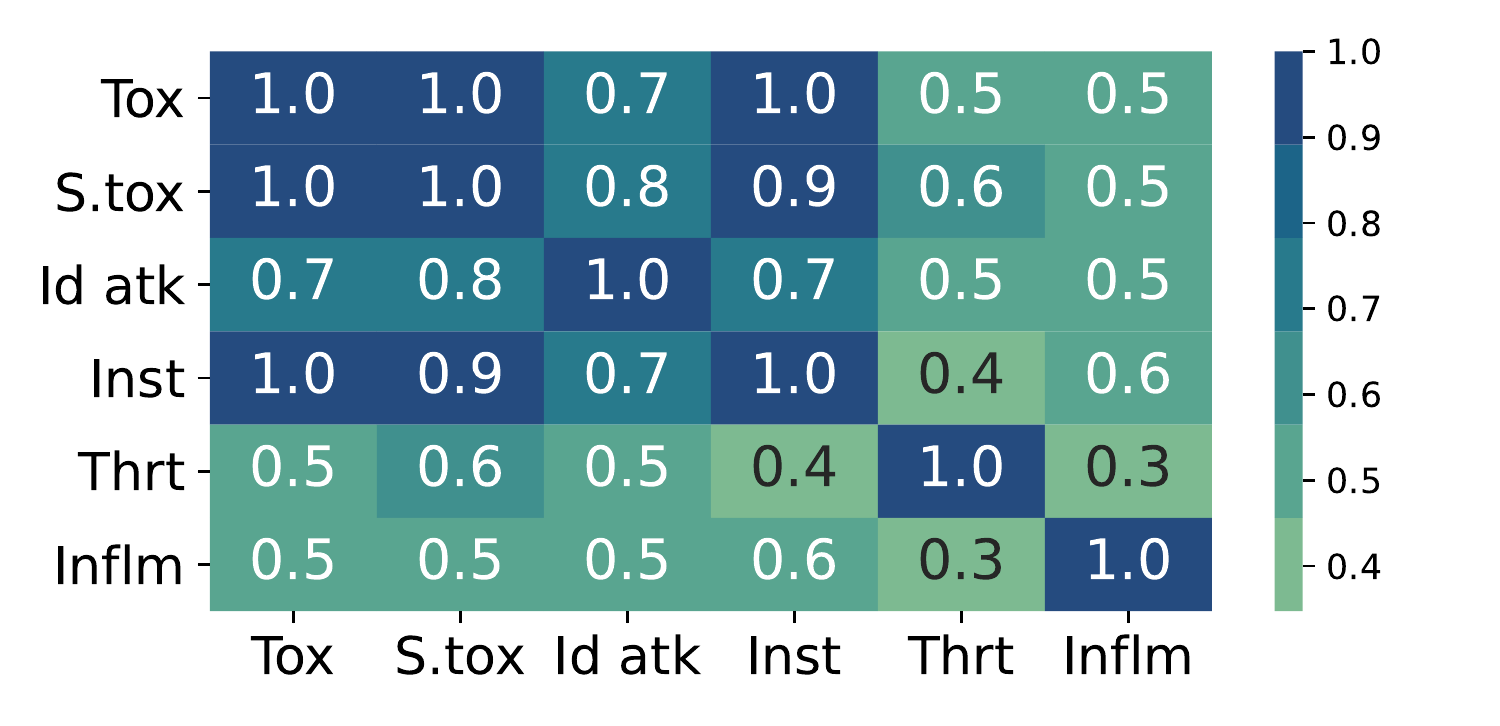}
            \vspace{-6mm}
            \caption{\small Pearson Correlation}
            \label{fig:cor_matrix}
    \end{subfigure}
    \vspace{-2mm}
    \caption{\small (\protect\subref{fig:perspective_scores_cdf}) Cumulative Distribution Function (CDF) of median Perspective API scores per profile, across all tweets; 
    (\protect\subref{fig:gini_index_cdf}) Gini Index calculated on all 6 Perspective scores per profile; and (\protect\subref{fig:cor_matrix}) Pearson pairwise correlation matrix across amongst all Perspective API scores.
    }
    \label{fig:characterize} 
    \vspace{-2mm}
\end{figure*}

\subsection{Crawling User Timelines}
\label{sec:user-timeline-crawl}
To provide broad coverage of themes, we merged 7 balanced seed datasets (in terms of toxicity) from various topic domains (Tab.\ref{tab:dataset_info}), and we estimate that this results in a dataset that closely reflects the Twitter community.
Based on Twitter's terms and conditions, Twitter User IDs (UIDs) and tweet content cannot be publicly shared. 
Consequently, our seed datasets contain only Tweet IDs (TIDs) and their respective annotation (Tab.\ref{tab:dataset_info}).
Thus, the first step was to query Twitter's API~\cite{twitterAPI} to recover the UID responsible for each TID.
Next, we queried the Twitter API to retrieve each UIDs' timeline.
Twitter API only permits the retrieval of 3,200 most recent tweets from a profile, whilst not the entire timeline, this still allows us to study a significant length of the historical record of each profile and its evolving behavior.
We were unable to retrieve tweets from banned and deleted profiles.
From the retrieved tweets, we extract relevant details such as the text, timestamp of tweet creation, hashtags, and URLs, shared within tweets. For this study, we only consider English tweets.

\subsection{Dataset Augmentation with Perspective API}
\label{sec:dataset-augmentation}
In the aforementioned seed datasets, a profile was labeled toxic or not based on mostly one or at most as few as 3 tweets.
However, it is unrealistic to assume that this label can be a representation of a profile's overall tweeting behavior. We needed a measure of misbehavior for the entire timelines of the 143K profiles we crawled. 
We obtain this quantitative measure of misbehavior across all tweets through Google's Perspective API~\cite{perspective}.
The Perspective API provides multiple Convolutional Neural Network (CNN) based models trained with GloVe word embeddings~\cite{pennington-etal-2014-glove} for the evaluation of misconduct in text.
This API offers 16 models of which 10 are considered experimental. Each model, for every given input text, yields a score from 0 to 1 representing the intensity of a type of misbehavior. 
We retrieve scores from the 6 production-ready Perspective API models~\cite{37635}:
\begin{itemize}[leftmargin=*]
    \item \emph{Toxicity}: Rude, disrespectful, or unreasonable comments, likely to make people leave a discussion.
    \item \emph{Severe Toxicity}: Comments are very likely to make users leave a discussion or give up sharing their opinion.
    \item \emph{Identity Attack}: Negative or hateful comments targeting someone's identity, ethnicity, sexual orientation, and other characteristics.
    \item \emph{Inflammatory}: Intended to provoke or inflame others.
    \item \emph{Insult}: Insulting or negative comments towards a person or a group of people.
    \item \emph{Threat}: Intentions to inflict pain, injury, or violence against an individual or group.
\end{itemize}
We query all 293M tweets from 143K profiles for the 6 perspective scores. The collective time for dataset collection and augmentation was about 5 months and the size of the augmented data is close to 2TB. We characterize the dataset in the following section.
\begin{table}[!t]
\centering
\resizebox{1.0\linewidth}{!}{
\begin{tabular}{l|r||l|r}
\toprule
\textbf{Recovered Profiles}  &  142,987 &\textbf{Total Profiles >10 Tweets}  & 138,533   \\ \hline
\textbf{Total Tweets}  &   293,401,160& \textbf{Avg Total Tweets per profile}  & 2,051 \\ \hline
\textbf{Unique Tweets}  &   230,283,810 & \textbf{Avg Unique Tweets per profile} & 1,610   \\ 
\midrule
\end{tabular}
}
\caption{\small Summary of Seed datasets (cf. Tab.~\ref{fig:data_collection_flowchart}) and recovered profiles.}
\label{tab:dataset_summary}
\end{table}

\subsection{Characterization of Augmented Dataset} 
\label{sec:3T-characterization}
To better understand the composition of our final dataset, we first inspect the Cumulative Distribution Function (CDF) of each Perspective score, across all tweets in all profile timelines (Fig.~\ref{fig:perspective_scores_cdf}).
We can observe that the median score of a tweet's score for any of the six misbehavior dimensions lies approximately in the range of 0.1 -- 0.2.
A steady rise in the curve in the low ranges of scores indicates that a majority of tweets do not strongly exhibit any specific form of misbehavior.

Additionally, the strongest signal for misbehavior is in the dimension of Inflammatory content.
A tail is observed with a small number of tweets acting as an exception to the rule, propagating a large amount of misbehavior (score $\rightarrow$ 1.0).

To measure a given profile's consistency in toxic behavior, we leverage the Gini Index over a profile's tweet perspective scores. The \textbf{\emph{Gini Index}} was originally proposed to measure the concentration of wealth~\cite{gini1912variabilita} within a population, but can equally be used to identify the extent of toxicity distribution among a profiles' tweets. 
A consistent set of scores (low or high) will produce a Gini Index closer to 0, whereas high variability scores produce a Gini Index approaching 1. 
We visualize the CDFs of the Gini index for all profiles across six dimensions of toxicity in Fig.~\ref{fig:gini_index_cdf}. We see that the median Gini is between 0.35 for Insult and 0.46 for Severe Toxicity. The majority of profiles have a Gini-Index in a range of 0.3-0.5 with a median of 0.4, which indicates that these profiles are not consistently toxic; however, we observe profiles in the lowest and highest range of Gini-Index 0-0.3 and > 0.6, pointing to profiles being constantly toxic. The Gini Index of Inflammatory scores is the lowest, implying that Inflammatory behavior is exhibited most consistently. Fig.~\ref{fig:cor_matrix} illustrates the correlation among all perspective scores, we note that toxic profiles are also likely to produce tweets that show identity attack and insult, and inflammation.

\subsubsection{\bf Takeaways:}
\begin{itemize}[leftmargin=*]
    \vspace{-1mm}
    \item With the range of median toxicity scores for all Twitter profiles between 0.14-0.16, we note that the majority of tweets on Twitter are not toxic.
    \item The majority of Twitter profiles have a low Gini index (0.4), thus they skew towards being consistently toxic across their tweets.
\end{itemize}
\subsection{Top 1\% Toxic Twitter Profiles}
\label{sec:top1p-profiles}
We identify and study the upper echelon of toxic profiles as determined by the average {``Toxicity''} Perspective score of their tweets. We sort all the 143K profiles based on the median toxicity scores calculated on the toxicity scores of all tweets in their respective timelines (at this point we remove all profiles with less than 10 tweets and consider the rest 138K profiles). 
As a final step, motivated by the fact that toxicity follows the Pareto effect on Twitter~\cite{ribeiro2018characterizing}, we skim the top 1\% of profiles as a sample of the most toxic Twitter profiles, we refer to these profiles (1,380) as \textbf{\emph{`Top 1\% toxic profiles'}}. Note that 80\% of the 1\% contribute 1000 or more tweets.
We shall compare toxic profiles with the remainder of the population (136,620, or 99\%), referred to as \textbf{\emph{`baseline profiles'}} in text. 
To contextualize the 1\% on toxicity scores, the 1\% profiles have a median toxicity of 0.40, while those in the baseline have a median of 0.15. Further, almost all tweets from the 1\% have a Toxicity larger than 0.35, whereas it lies at <3\% for the baseline.

\section{Tweet frequency and pattern analysis}
\label{sec:prolificacy_analysis}
Impactful Twitter profiles post a significant number of tweets over time. In this section, we shall investigate the number of total and  unique (tweets with the exact same content were removed) tweets, and the percentage of unique tweets posted by toxic 1\% and baseline profiles. We also consider the tweeting pattern as a measurable trait. It reveals the longitudinal nature of a profile's posting behavior.

\subsection{Tweet Frequency}
\label{subsec:tweeting_frequency_analysis}
\subsubsection{Are toxic profiles prolific?}
\begin{figure}[t]
\centering
\subfloat[Tweets count.]{
 \includegraphics[width=0.49\columnwidth]{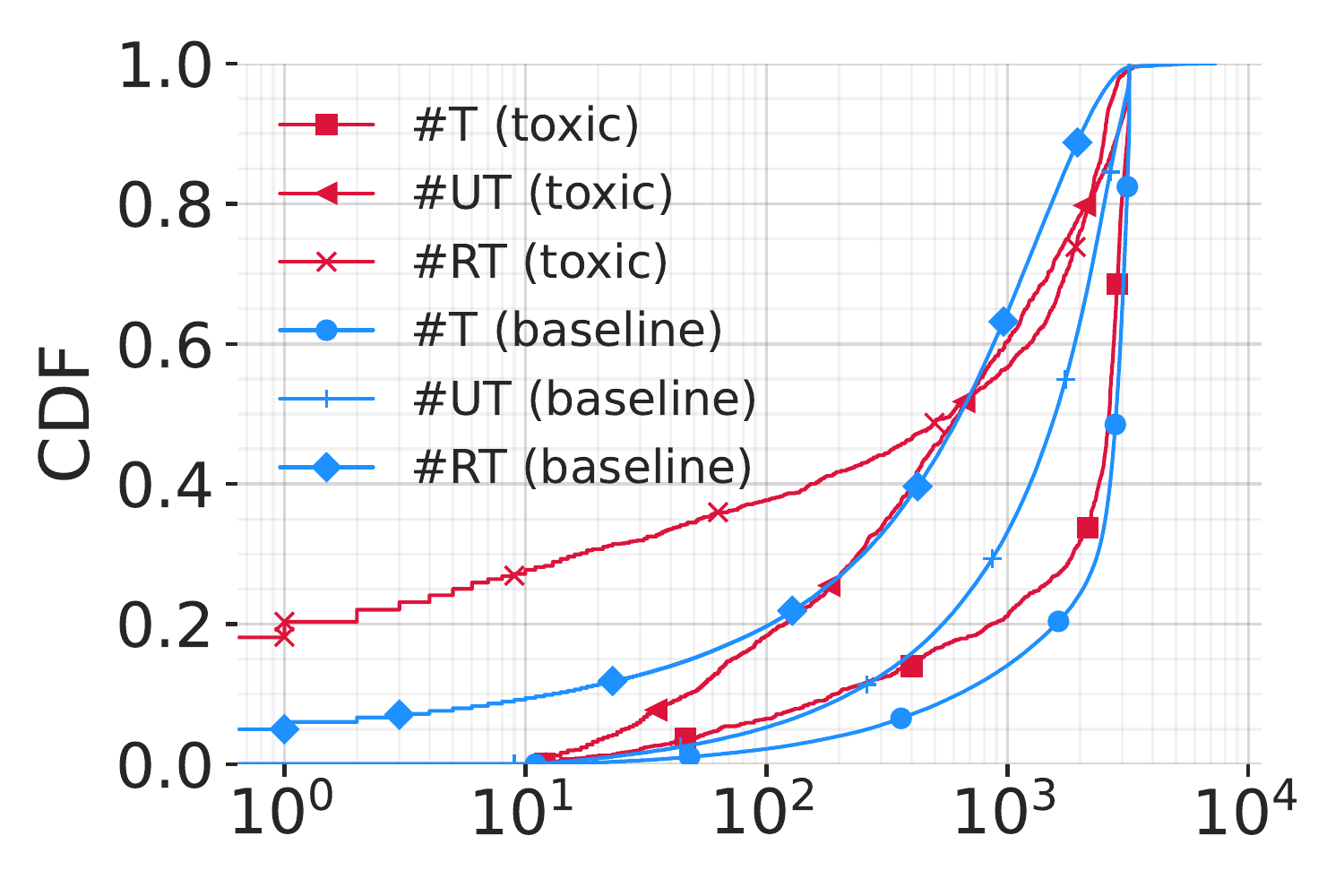}\label{fig:tweet_stats}
}
\subfloat[Percentage of unique tweets.]{
\includegraphics[width=0.48\columnwidth]{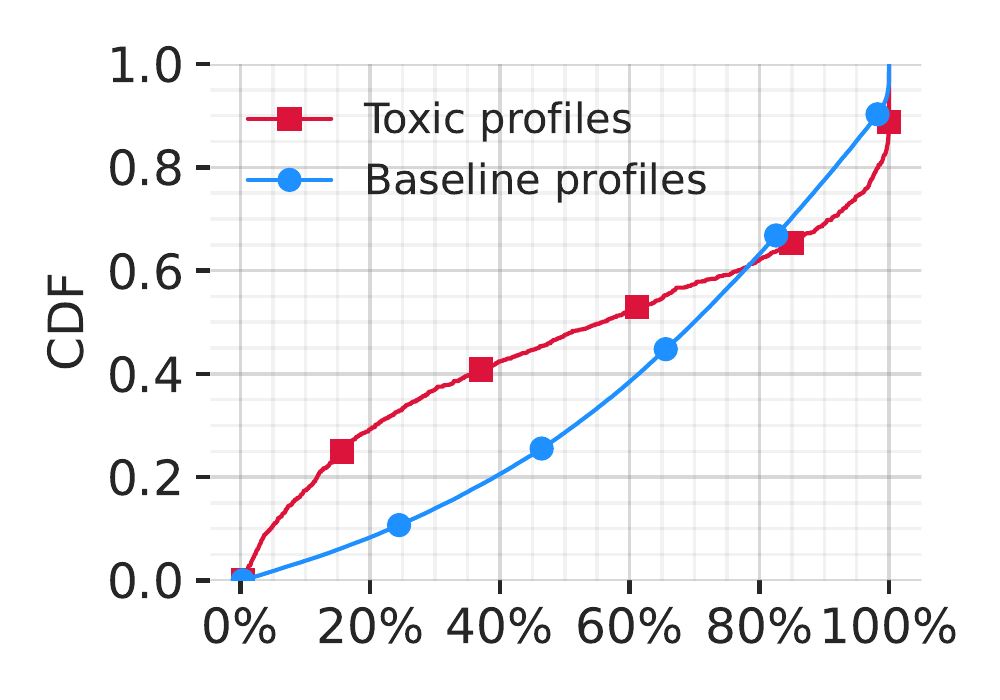}\label{fig:tweet_stats_perc_uni}
}
\vspace{-2mm}
\caption{\small Tweeting frequency of toxic and baseline profiles (\S\ref{subsec:tweeting_frequency_analysis}). In \protect\subref{fig:tweet_stats}, T, RT, and UT, respectively, represent the number of Tweets, retweets (RT), and unique tweets per profile.}
    \vspace{-4mm}
\end{figure}

In order to uncover the answer, we first note the total number of tweets of toxic 1\% and baseline profiles. 
Unique tweets (repetitions removed) were further counted to observe whether profiles repeatedly repost the same tweet. 
We present a CDF with the number of total and unique tweets in Fig.~\ref{fig:tweet_stats}. From this figure, we observe that 80\% of toxic profiles in our dataset post more than 1,000, as compared to 82\% in baseline profiles. It is interesting to observe that half of the toxic profiles at most tweeted 500 unique tweets and half of the baseline profiles at most posted 1500 unique tweets showing that toxic profiles almost tweet equal to baseline but post fewer original tweets. We note that base 40\% of our toxic and 20\% of baseline profiles post retweets less than 100 times.
Fig.~\ref{fig:tweet_stats_perc_uni} details the repetitive behavior of our profiles, it is evident that half of the toxic profiles in our dataset posted at most 55\% unique tweets (no repeats) and this is true for only a quarter of baseline profiles. Interestingly, a larger proportion of profiles in the toxic set occupy lower percentages of unique tweets, before crossing over with the baseline at 77\% unique tweets. At the higher percentages of unique tweets, the baseline increases gradually, whilst the bulk of the remaining toxic profiles have near 100\% unique tweets, with 25\% of toxic 1\% profiles posting more than 95\% unique tweets compared to only 15\% baseline profiles and 15\% of toxic 1\% profiles with no repetition vs. 7\%. Thus there is the occurrence of toxic profiles that repeatedly re-post the same toxic message, and profiles with individually crafted tweets containing toxicity. We report that the median toxicity of re-posted tweets is 0.47 for the toxic and 0.32 for the baseline.

\subsubsection{\bf Takeaway:}
Toxic profiles are comparable to general Twitter profiles in the total number of posted tweets but they retweet less than the baseline profiles. Notably, about 20\% of toxic profiles have a much higher proportion of unique tweets than the baseline.

\begin{figure*}[!t]
\centering
    \begin{subfigure}[t]{0.39\linewidth}
    \includegraphics[width=\linewidth]{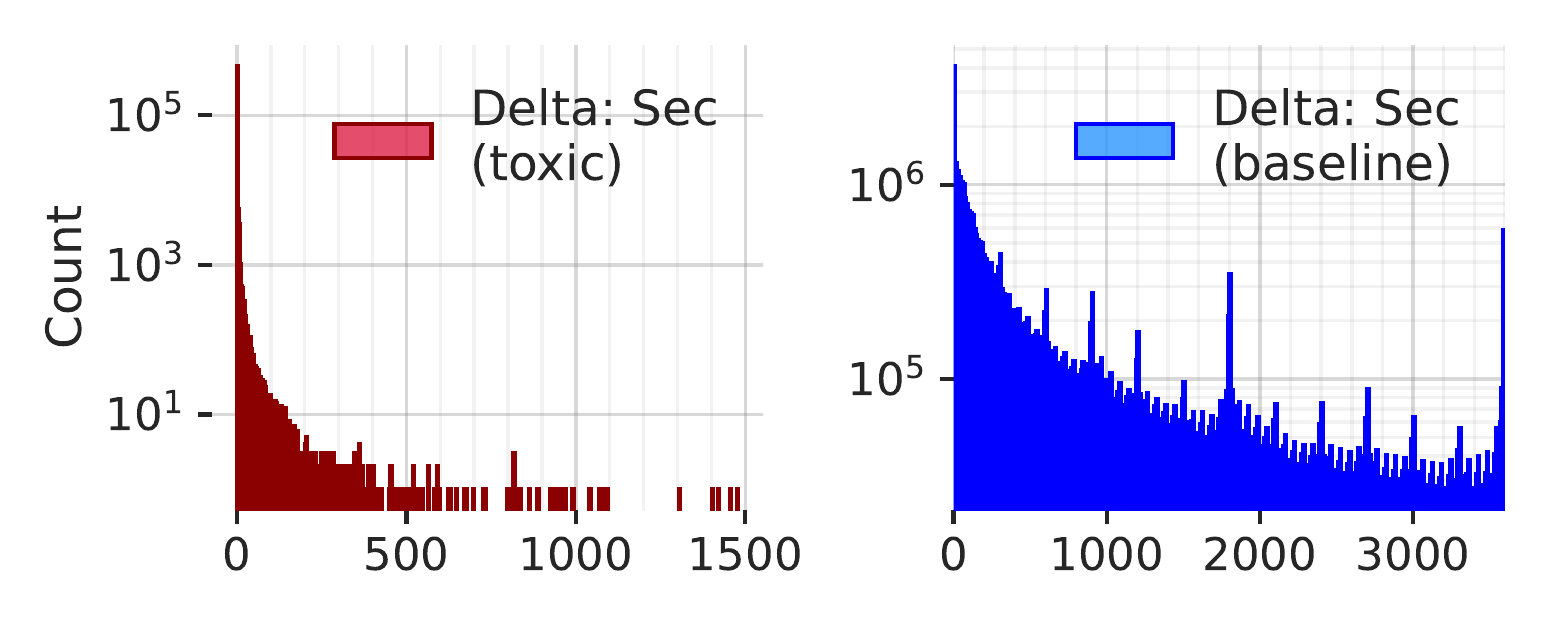}
    \vspace{-8mm}
    \caption{{\small Seconds between consecutive tweets}}
    \label{fig:delta_combined_sec}
    \end{subfigure}
    \begin{subfigure}[t]{0.39\linewidth}
    \includegraphics[width=\linewidth]{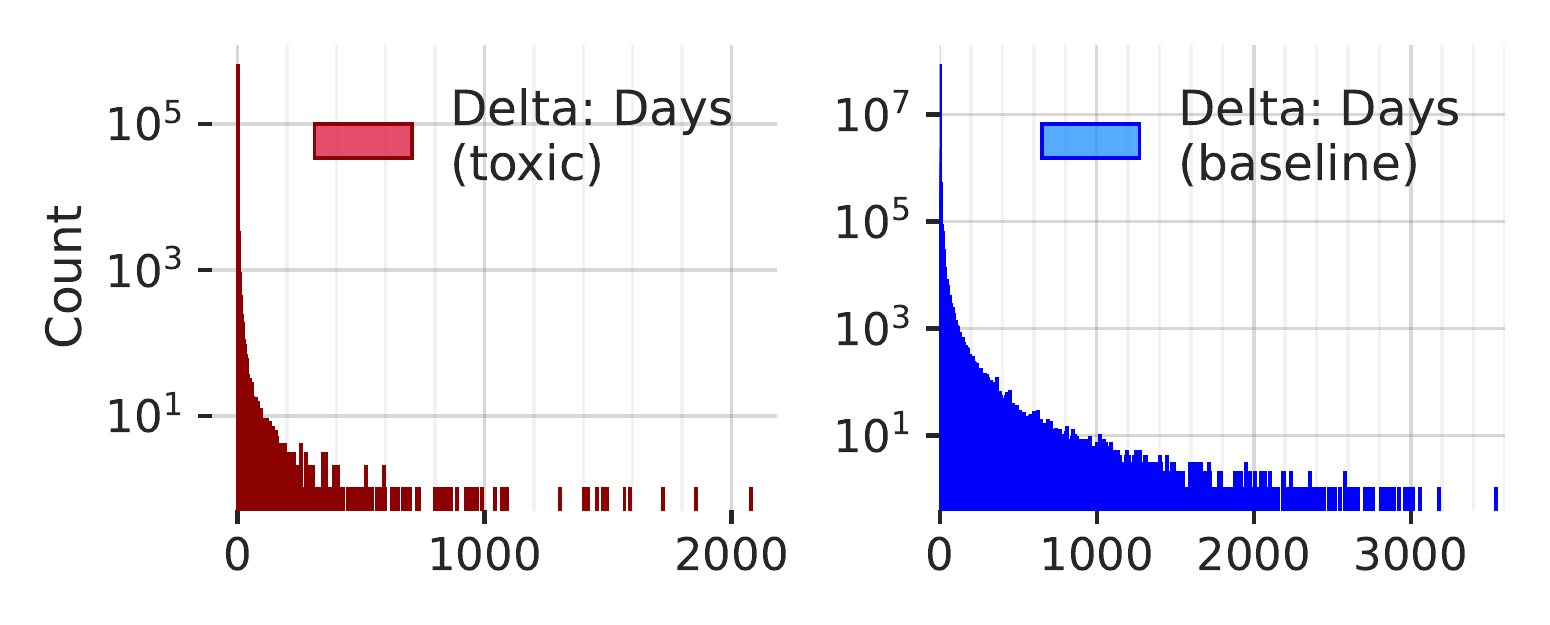}
    \vspace{-8mm}
    \caption{{\small Days between consecutive tweets}}
    \label{fig:delta_combined_days}
    \end{subfigure}
    \begin{subfigure}[t]{0.19\linewidth}
    \includegraphics[width=\linewidth]{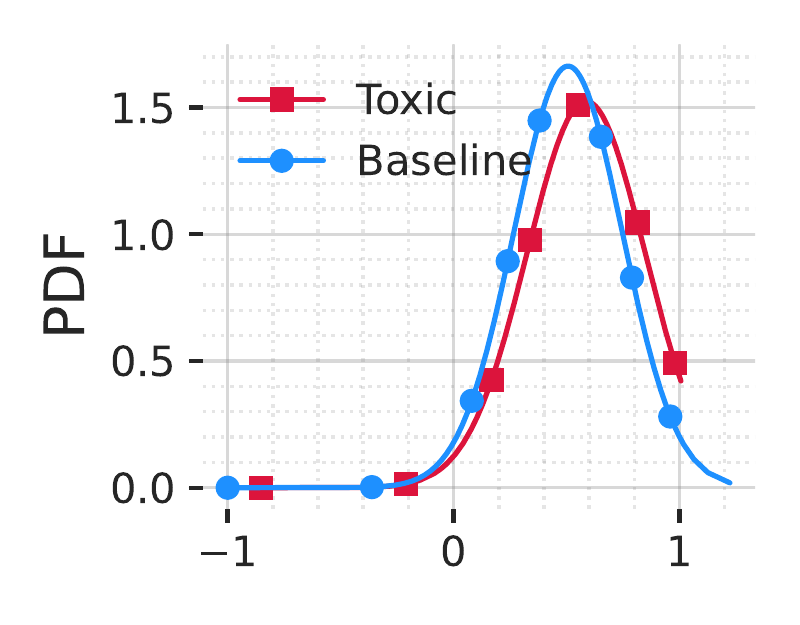}
    \vspace{-8mm}
    \caption{{\small Burstiness Score}}
    \label{fig:burstiness_scores}
    \end{subfigure}
    \vspace{-0.1cm}
    \caption{\small Tweeting pattern of toxic and baseline profiles in time between sequential tweets, and burstiness in time (cf. \S\ref{subsec:tweeting_pattern}).}
    \vspace{-0.5mm}
\end{figure*}
\subsection{Tweet Pattern}
\label{subsec:tweeting_pattern}
\subsubsection{Do toxic profiles follow any particular tweeting pattern?}
To capture a profile's tweeting manner, we consider the \textbf{\emph{Time delta}} i.e., the time between sequential or consecutive tweets (time noted from tweet timestamp) for each profile in our dataset. We consider histograms of time deltas in seconds and days in Fig.~\ref{fig:delta_combined_sec} and \ref{fig:delta_combined_days} respectively. 
For both toxic and baseline groups, the most populated period of time between tweets is within a few seconds. It is interesting to observe the occurrence of periodic behavior within the baseline profiles, indicating the presence of automation, despite these profiles not being overly toxic.
On the other hand, toxic profiles do not appear to post regularly at fixed intervals. There is a clear skew to the shorter time deltas in toxic profiles compared to the baseline.

\subsubsection{Is there consistency in the temporal tweeting pattern of the toxic profiles?}
We now explore if the time differences between the tweets that we have observed are distributed consistently within a profile's timeline, or if profiles `activate' briefly for bursts of activity and then go dormant.

To this end, we employ a normalized measure of burstiness to compare the tweeting behavior of toxic and baseline profiles. \textbf{{\emph{Burstiness}}}~\cite{kim2016measuring} is a quantification of inter-event time distribution from a given event sequence, that is, the distribution of time deltas between consecutive events. The 
\emph{Burstiness Score}
$
    B=\frac{\sigma-\mu}{\sigma+\mu}=\frac{r-1}{r+1},
    \label{eq:burstiness}
$
where $r=\sigma/\mu$ is the
coefficient of variation and $\sigma, \mu$ denote the standard deviation and mean of inter-event time distribution respectively. 
$B$ ranges continuously between $-1$ and $1$; regular time series (near constant inter-event times) would have scores closer to $B=-1$, $B=0$ is a random sequence, and $B=1$ is an extremely bursty time series (as $\sigma\rightarrow\inf$ for finite $\mu$). 
It is known that burstiness is dependent on the length of the time series, and since the number of events (tweets) differs among profiles, we adopt \textbf{\emph{Normalized Burstiness}}~\cite{kim2016measuring}, which removes this dependency, to facilitate direct comparison among profiles.
$B(n,r) =\frac{\sqrt{n+1} r - \sqrt{n-1}} {(\sqrt{n+1}-2) r+\sqrt{n-1}}$,
note that $B(n,r)$ can take values greater than one and less than -1. 
Fig.~\ref{fig:burstiness_scores} provides a probability density function or PDF of burstiness per profile. We observe that toxic profiles skew towards being more bursty with a curve peak at 0.6 as compared to 0.5 for the baseline Twitter profile. A two-sample t-test 
yields a p-value of $4.26\times10^{-26}$, considerably less than $(\alpha=0.05)$ to indicate that the distribution of Burstiness is significantly different between the toxic and baseline profile groups. 

From this, we conclude that the toxic profiles are more irregular (and more bursty) in their tweeting behavior than the baseline profiles in general.
\subsubsection{\bf Takeaways:}
\begin{itemize}[leftmargin=*]
    
    \item Toxic profiles can be long-lived accounts with activity gaps of 8-9 years, and they are more likely to tweet in quick succession with minimal activity intervals.
    \item Toxic profiles do not appear to tweet at regular fixed intervals (a sign of automation), a behavior observed in the baseline profiles.
    \item Generally parallels can be drawn between the temporal behavior of baseline and toxic profiles, however toxic profiles skew to favor shorter intervals between tweets and are more bursty.
\end{itemize}

\begin{figure*}[!ht]
    \centering
        \begin{subfigure}[t]{0.135\linewidth}
        \centering
            \includegraphics[width=\textwidth]{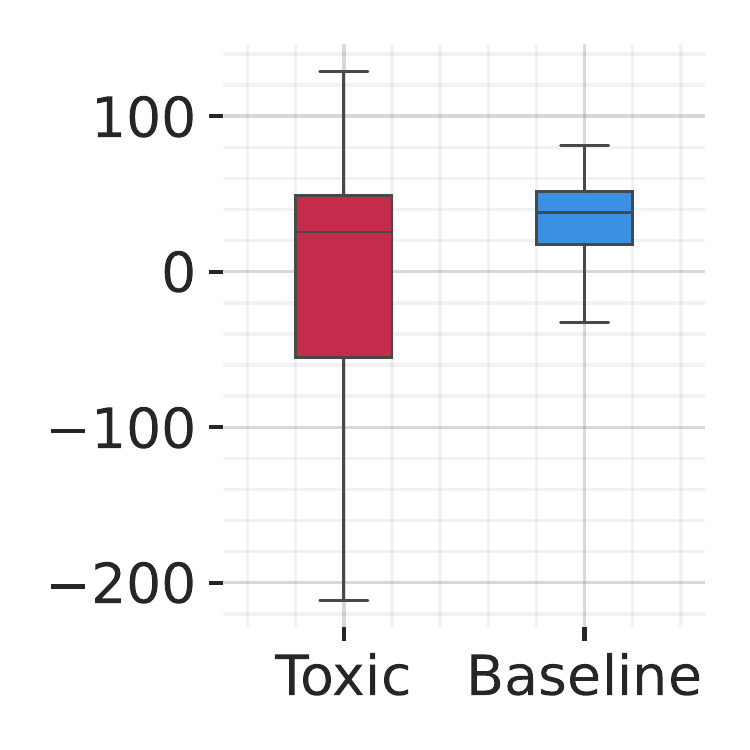}
            \vspace{-3mm}
            \caption{\small Flesch Kincaid\newline reading ease score}
            \label{fig:fkr_score}
    \end{subfigure}
    \begin{subfigure}[t]{0.135\linewidth}
            \centering
            \includegraphics[width=\textwidth]{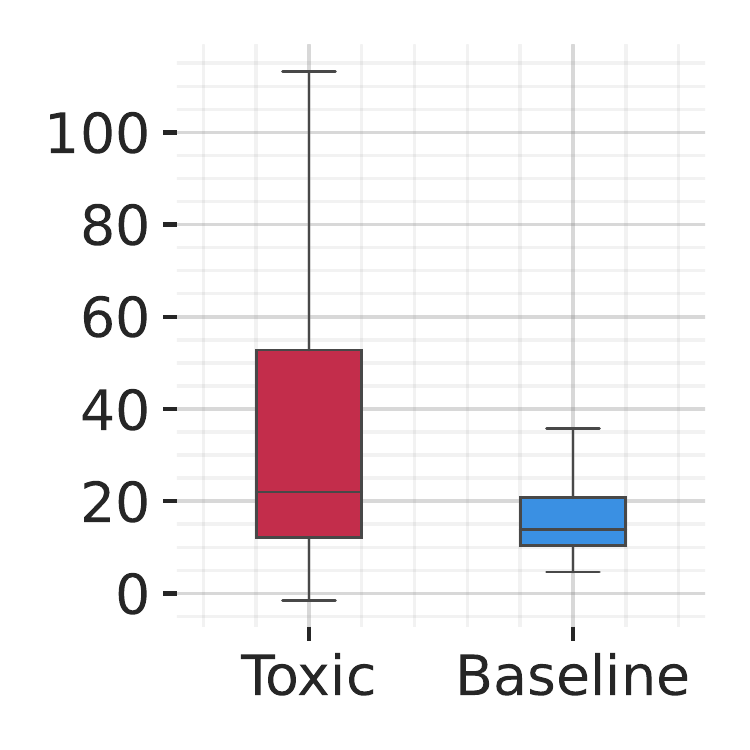}
            \vspace{-3mm}
            \caption{\small Flesch Kincaid\newline difficulty score}
            \label{fig:fk_score}
    \end{subfigure}
    \begin{subfigure}[t]{0.135\linewidth}
            \centering
            \includegraphics[width=\textwidth]{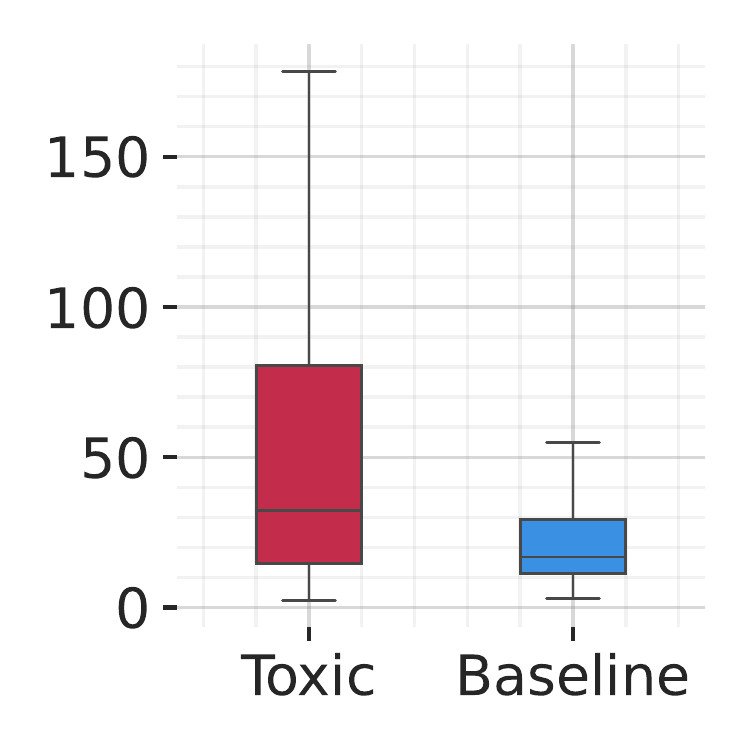}
            \vspace{-3mm}
            \caption{\small Linsear write \newline score}
            \label{fig:lin_write_score}
    \end{subfigure}
    \begin{subfigure}[t]{0.135\linewidth}
            \centering
            \includegraphics[width=\textwidth]{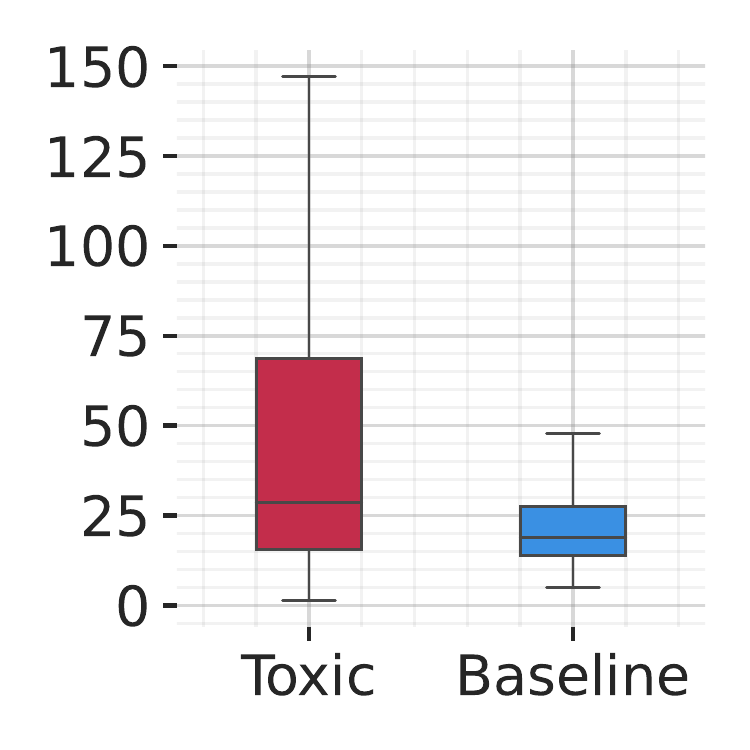}
            \vspace{-3mm}
            \caption{\small ARI \newline score}
            \label{fig:auto_read_score}
    \end{subfigure}
    \begin{subfigure}[t]{0.135\linewidth}
            \centering
            \includegraphics[width=\textwidth]{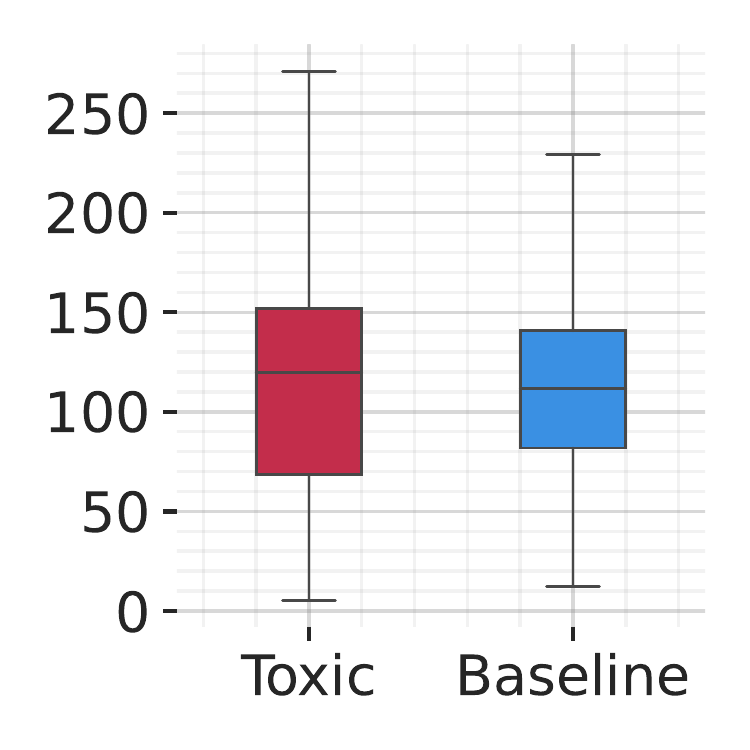}
            \vspace{-3mm}
            \caption{\small Lexical \newline diversity score}
            \label{fig:lexical_diversity}
    \end{subfigure}
     \begin{subfigure}[t]{0.135\linewidth}
            \centering
            \includegraphics[width=\textwidth]{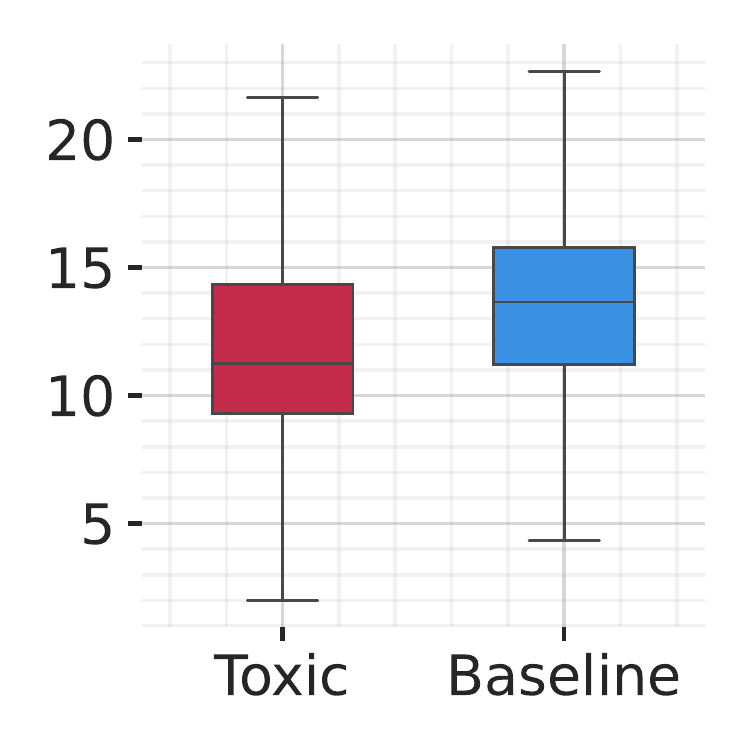}
            \vspace{-3mm}
            \caption{\small \# Average \newline tweet words }
            \label{fig:tweet_words}
    \end{subfigure}
    \begin{subfigure}[t]{0.135\linewidth}
            \centering
            \includegraphics[width=\textwidth]{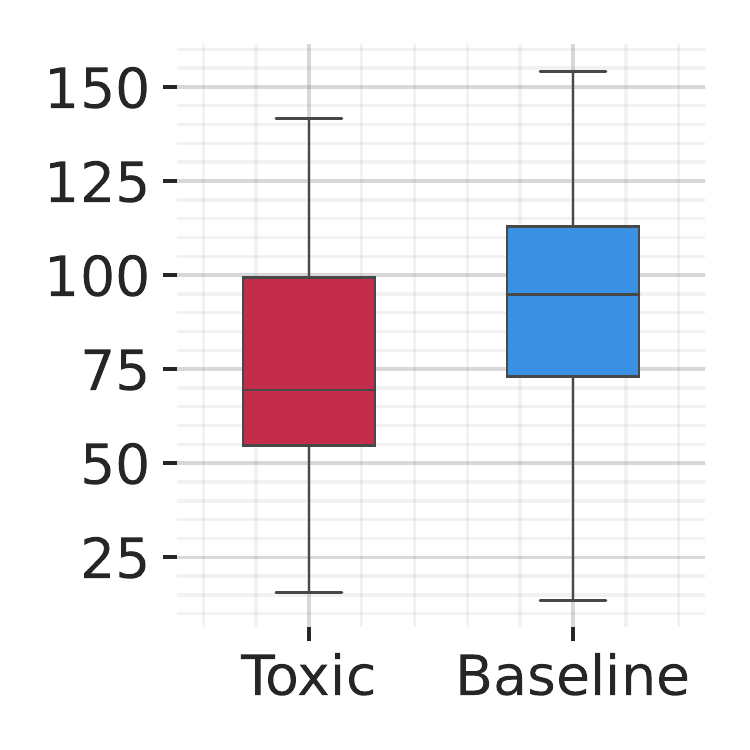}
            \vspace{-3mm}
            \caption{\small \# Average \newline tweet char}
            \label{fig:tweet_char}
    \end{subfigure}
    \vspace{-2mm}
    \caption{\small Lexical analysis of toxic and baseline profiles' tweets (cf. \S\ref{subsec:lexical_analysis} for details) in our dataset.}
    \label{fig:lexical_analysis}
    \vspace{-2mm}
\end{figure*}

\begin{figure*}[!ht]
    \centering
    \begin{subfigure}[t]{0.4\linewidth}
            \includegraphics[width=\textwidth]{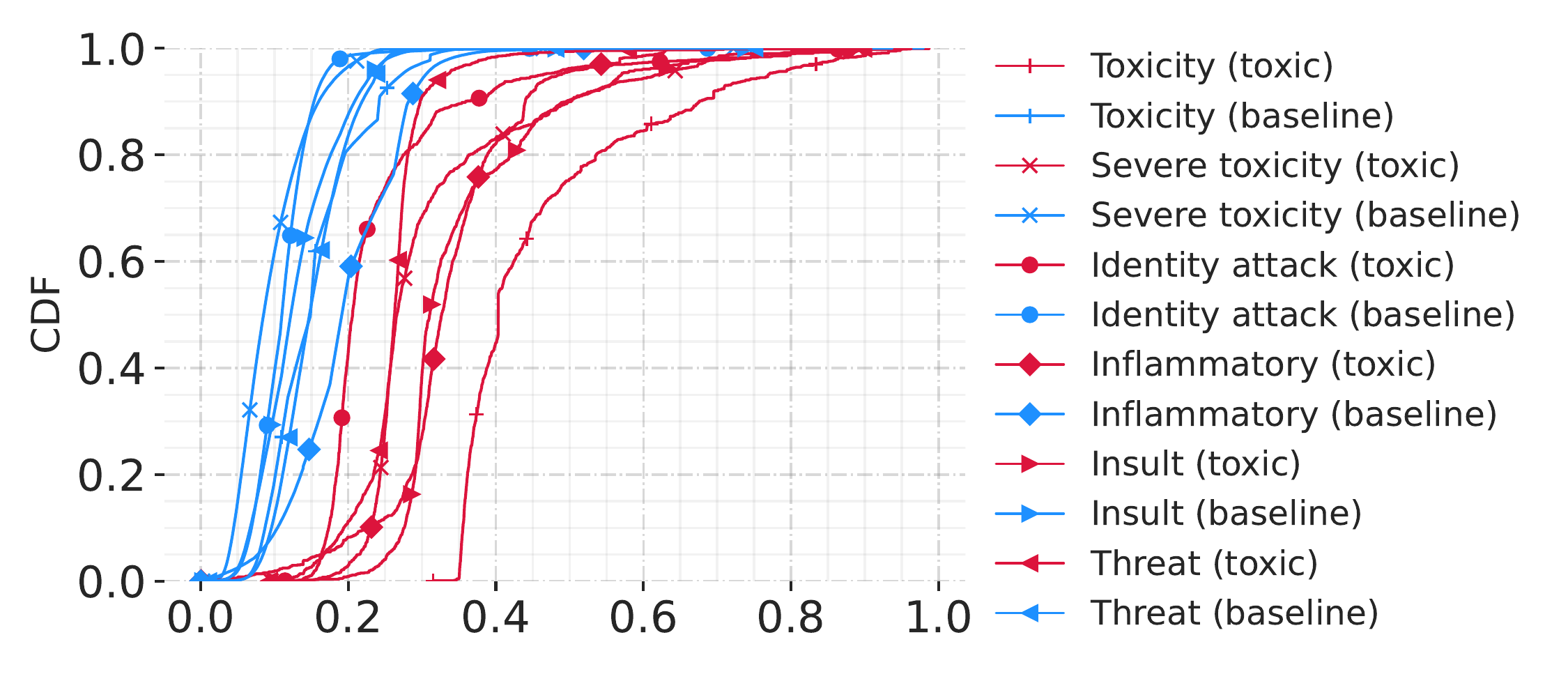}
            \vspace{-8mm}
            \caption{{\small Median Perspective scores}}
            \label{fig:perspective_scores}
    \end{subfigure}
    \qquad\qquad
    \begin{subfigure}[t]{0.4\linewidth}
            \includegraphics[width=\textwidth]{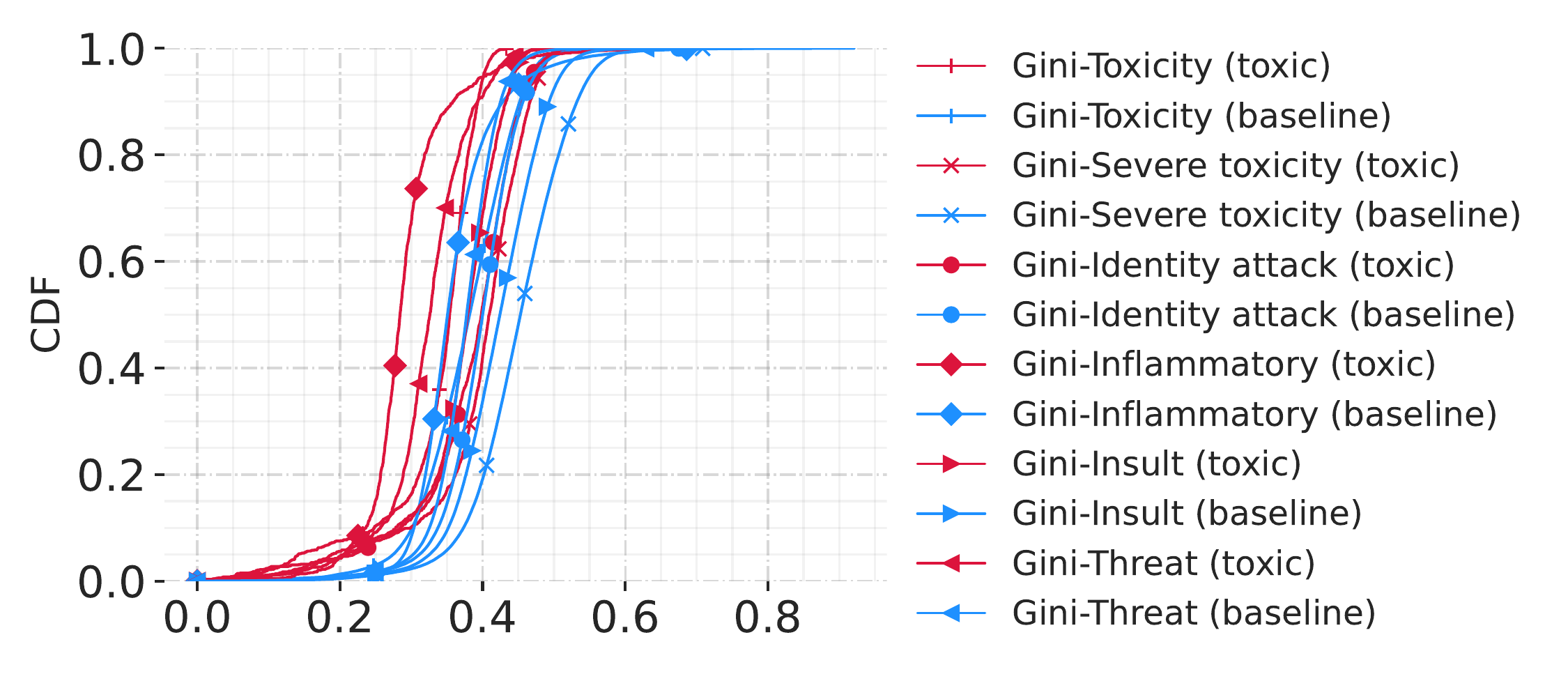}
            \vspace{-8mm}
            \caption{{\small Gini index of Perspective scores}}
            \label{fig:gini_scores}
    \end{subfigure}
    \vspace{-2mm}
    \caption{\small Median (\protect\subref{fig:perspective_scores}) and Gini index (\protect\subref{fig:gini_scores}) of Perspective API scores (cf. \S\ref{sec:toxicity_analysis} for details).
    }
    \vspace{-4mm}
\end{figure*}

\section{Content analysis}
\label{sec:content_analysis}
The nature of a profile's tweets is determined by the actual text and, also by attached auxiliary content such as URLs and hashtags.
We perform a content analysis of timeline tweets with respect to each toxic or baseline profile. Specifically, we analyze the quality of the tweet's text (\S\ref{subsec:lexical_analysis}), toxicity level in the text (\S\ref{sec:toxicity_analysis}), and additional tweet attributes like URLs (\S\ref{subsec:url_analysis}), and hashtags (\S\ref{subsec:hashtag_analysis}). 

\subsection{Tweet lexicon} 
\label{subsec:lexical_analysis}
The way a tweet is constructed tells the degree of authority of the author and potential target audience. Thus, we now analyze the text within our profiles' tweets for length, grammatical correctness, and semantic correctness. 
\subsubsection{Do toxic profiles share verbose tweets?}
\label{subsectio:verbosity_analysis}
We question, how much of the character allowance in a tweet is utilized by our toxic profiles. 
To this end, we parse each tweet to extract the number of words and characters for both toxic and baseline profiles. The boxplots in Fig.~\ref{fig:tweet_words} and~\ref{fig:tweet_char} display the distribution of the average number of words and characters in tweets. 
We observe that toxic profiles post shorter tweets with fewer words than baseline Twitter profiles with an average of 11 words and 70 characters.
\subsubsection{Do toxic profiles share legible and easy-to-read tweets? }
With the tweet text in hand, we measure the Flesch Score~\cite{flesch1948new} (ease and difficulty), Linsear write scores, Automated Readability Index~\cite{senter1967automated}, and Lexical Diversity of toxic and baseline profiles. 

The \textbf{\emph{Flesch score}} indicates how difficult or easy it is to read the text~\cite{flesch}, and is computed as:
$
    206.835-1.015\times(\frac{total words}{total sentences})-84.6\times(\frac{total syllables}{total words})
$.
\textbf{\emph{Linsear write score}} measures the length of words in the number of syllables and divides this score ``r'' by the number of total sentences~\cite{mccannon2019readability}. If (r > 20, Lw = $\frac{r}{2}$) and if (r$\ge$20, Lw = $\frac{r}{2-1}$). 
\textbf{\emph{Automated Readability Index (ARI}} estimates the comprehensibility of a text corpus and is computed as (4.71$\times$average word length)+(0.5$\times$average sentence length)-21.43~\cite{ARI}. 
\textbf{\emph{Lexical diversity}}, defined as the ratio of a number of unique words to the total number of words, reveals noticeable repetitions of distinct words~\cite{doi:10.1080/15434303.2020.1844205}.
Higher values of the ARI, Flesch scores and lexical diversity of a given text indicate increased comprehensiveness, improved readability, and range and variety of vocabulary. 
A given text with a high Linsear write score generally includes words with more syllables and/or is written with richer language.

Fig.~\ref{fig:lexical_analysis} provides a summary of the 3 metrics.
In comparison to baseline profiles, toxic profiles share more legible and readable tweets. Toxic profiles use richer and more profound vocabulary, which as explained above is a predictable use of language and a sign of good writing style.
\subsubsection{\bf Takeaway:}
Top toxic 1\% profiles share shorter tweets written in more understandable language than baseline.

\subsection{Tweet Toxicity}
\label{sec:toxicity_analysis}
\subsubsection{What type of misbehavior is common in toxic profiles?}  
The 6 scores from the Perspective API provide granular insight into the specific types of misbehavior exhibited by a profile. 
We plot the median scores of Toxicity, Severe Toxicity, Identity Attack, Inflammatory, and Insult per profile as a CDF in Fig.~\ref{fig:perspective_scores}. We observe that toxic profiles on all 6 dimensions of misbehavior exceed that of general Twitter profiles. We note that beyond ``Toxicity'', tweets high in ``Inflammatory'' and ``Insult'' are the next most prevalent within our toxic profiles. Interesting to note that Inflammatory is comparatively less prominent (after toxicity) among toxic profiles than baseline profiles. On the other hand, the lowest score for toxic profiles is ``Identity Attack'', This would be consistent with Twitter policy, which states that racist tweets are not tolerated~\cite{Twitter_conduct_policy}.
Figure~\ref{fig:gini_scores} is a CDF of the Gini index calculated on all 6 toxicity scores to gauge the consistency of misbehavior amongst a profile's tweets. We observe that toxic profiles in comparison to baseline profiles have lower Gini scores, implying the top 1\% toxic profiles exhibit misbehavior relatively consistently compared to the baseline.
\subsubsection{\bf Takeaway:}
The most common forms of toxicity among the Toxic 1\% profiles are ``Inflammatory'' and ``Insult'', with ``Identity Attack'' the least common.

\begin{figure}[!htb]
\centering
    \begin{subfigure}[t]{0.46\linewidth}
        \includegraphics[width=0.95\linewidth]{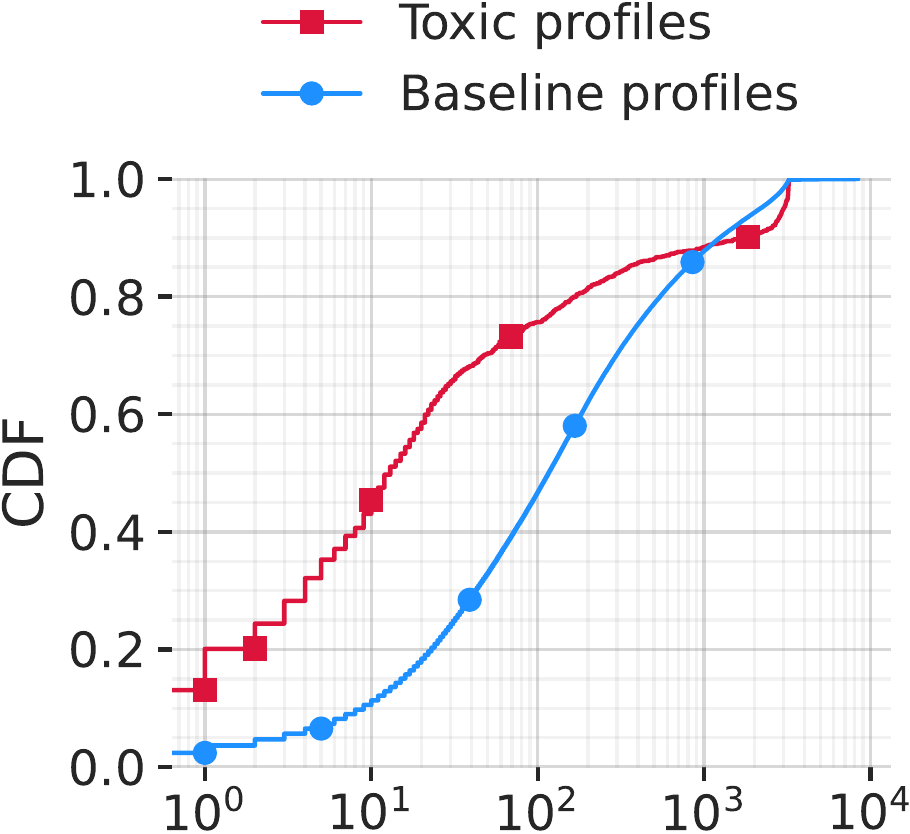}
        \vspace{-1mm}
        \caption{\#URLs}
        \label{fig:no_total_urls}
    \end{subfigure}
    \hfill
    \begin{subfigure}[t]{0.46\linewidth}
        \includegraphics[width=0.95\linewidth]{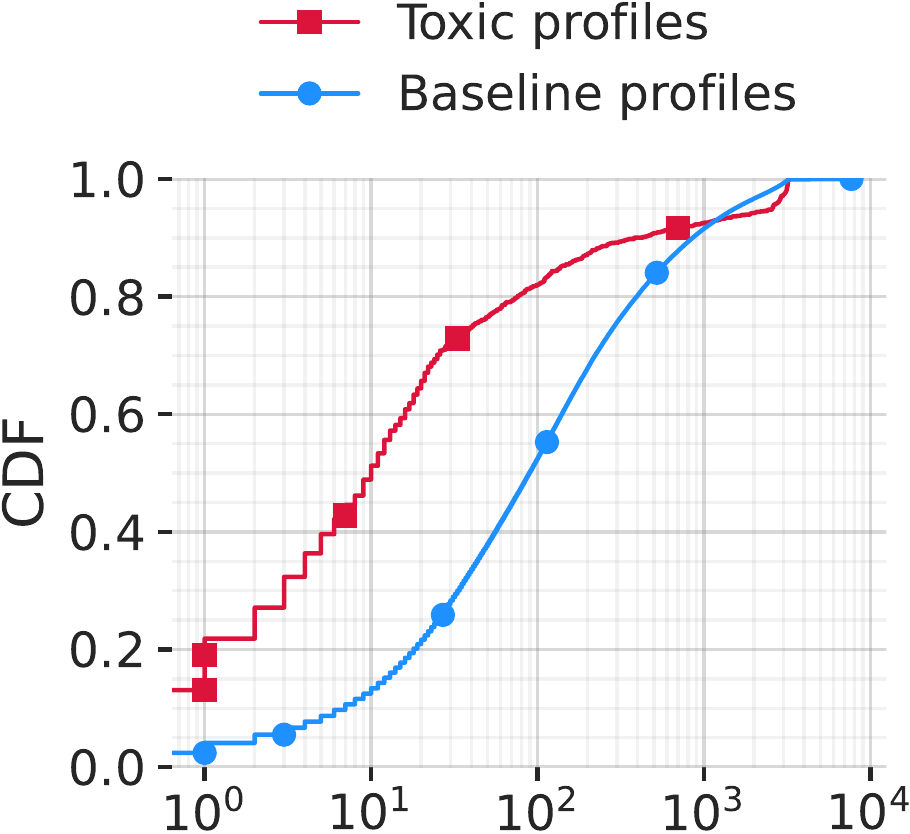}
        \vspace{-1mm}
        \caption{\# Unique URLs}
        \label{fig:no_unique_urls}
    \end{subfigure}

    \begin{subfigure}[t]{0.46\linewidth}            \includegraphics[width=0.95\linewidth]{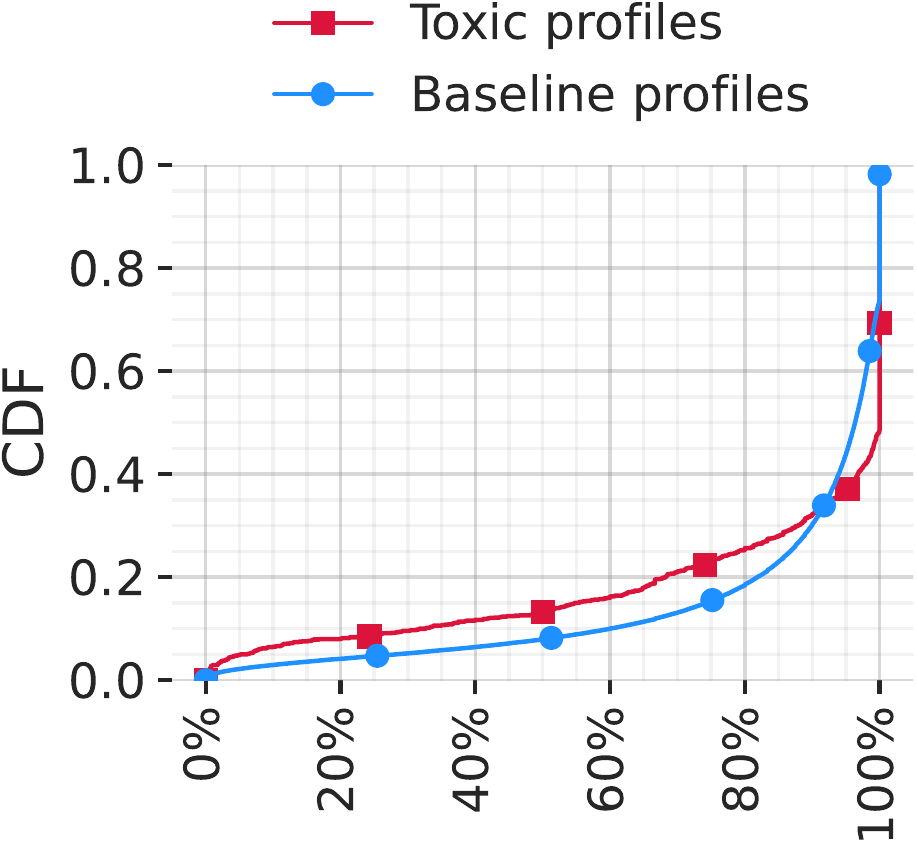}
        \vspace{-1mm}
        \caption{{\% Unique URLs }}
        \label{fig:p_unique_urls_combined_cdf}
    \end{subfigure}
    \hfill
    \begin{subfigure}[t]{0.46\linewidth}
        \includegraphics[width=0.95\linewidth]{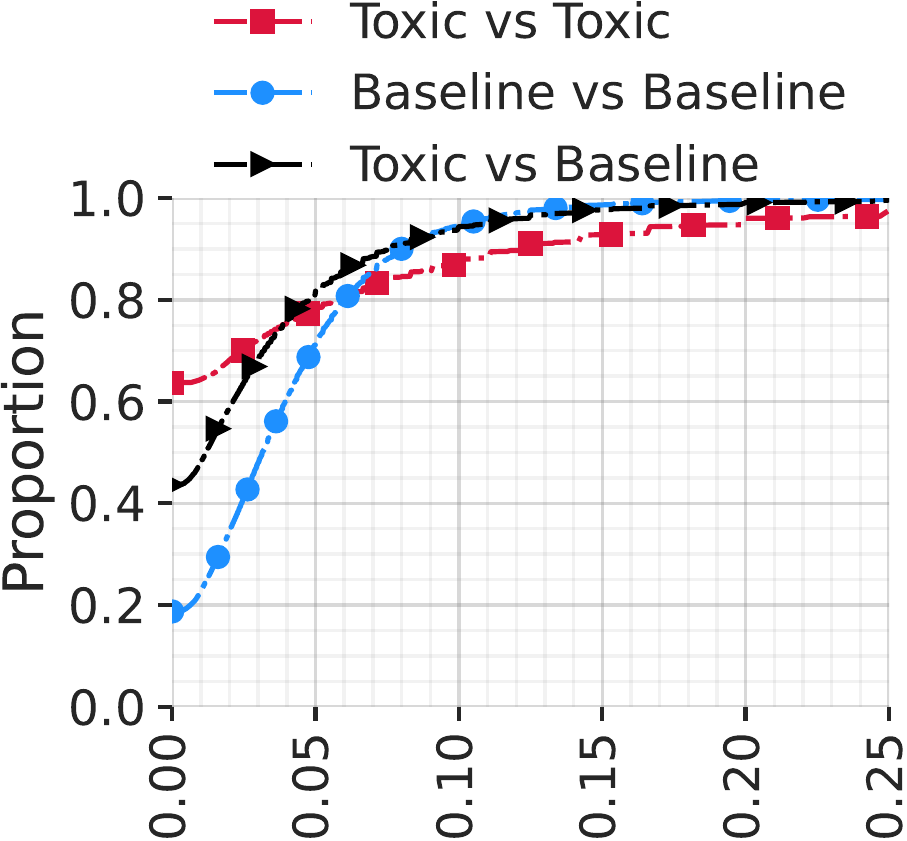}
        \vspace{-1mm}
        \caption{{Jaccard Similarity (SLDs)}}
        \label{fig:jaccard_similarity}
    \end{subfigure}
    
    \vspace{-1mm}
    \caption{\small Profile level URL analysis (\S\ref{subsec:url_analysis}).}
    \vspace{-2mm}
\end{figure}

\subsection{URLs}
\label{subsec:url_analysis}
A shared URL is an indication that a profile seeks to point a reader to a resource external to the Twitter platform, either as a corroborative source of validation or for further reading about their tweet's subject matter. On the other hand, The repetition of a shared URL and posting URLs related to one subject (category) shows how much a profile emphasizes a topic.
From our 1,380 toxic profiles and 136,620 baseline profiles, we detect and extract a total of 57,725,668 URLs and 43,916,037 unique URLs in total. 
\subsubsection{How frequently do toxic profiles share URLs as part of their tweet text?}

To answer this question, we count total number of URLs and also note the number of URLs without repetition (we inspect the full length of the URLs, extracted from the tweet's metadata and we did not rely on the shortened version used in the tweet text). 
Fig.~\ref{fig:no_total_urls} illustrates a CDF on the total number of URLs per profile for both groups. On the low end of the figure, it is clear that baseline profiles engage more with sharing URLs than toxic profiles. We observe that 45\% of the toxic profiles shared 10 or fewer URLs in their tweets, compared to only 10\% of baseline profiles. 
On the other extreme, approximately 10\% of both profile groups are heavy URL hitters with more than 1,000 URLs in total, and 5\% of toxic profiles posted ~3,200 URLs compared to nearly no baseline profiles. 3,200 corresponds to the maximum number of tweets obtainable from a single profile. We note that 3.3\% baseline profiles shared ~6-8K URLs, these profiles shared multiple URLs per tweet in short form and on average shared 3 URLs per tweet --- these profiles were predominantly news services can also comment from \S\ref{subsec:tweeting_pattern} that there are profiles in baseline which are persistent with regular tweeting pattern, which might indicate these are bots.  

Fig.~\ref{fig:no_unique_urls} presents the unique number of URLs per profile (i.e.:~not counting repetitions). Half of the toxic profiles shared at most 0-10 unique URLs and half of the baseline profiles shared at most 0-95 unique URLs. It is interesting to observe that 22\% of toxic profiles used a singular unique URL and 13\% did not post any URLs at all.
Fig.~\ref{fig:p_unique_urls_combined_cdf} shows the proportion, per profile, of URLs that are repetitions among toxic and baseline profiles. We see that around 48\% of toxic profiles do not repeat URLs at all, compared to only 27\% of baseline profiles (right-hand side of the plot). 
In contrast, looking at profiles that repeat URLs the most, the top 33\% of toxic profiles (CDF values 0.0 to 0.33) have substantially more repetition than the corresponding group of baseline profiles. 
We note that toxic profiles in general share a lower percentage of unique URLs in their tweets. 
Fig.~\ref{fig:jaccard_similarity} shows us the Jaccard similarity of URLs amongst and between toxic and baseline groups. We observe that toxic profiles share URLs of the same nature.
A large proportion of toxic profiles (63.7\%) have no URL similarity with one another, in comparison to between baseline profiles (18.7\%), this could be the result of the toxic profiles operating independently, or with uniquely crafted/tracking URLs. The remaining 20\% of toxic profiles however do have a heightened shared URL similarity compared to the baseline, indicating the existence of coordination.

\begin{figure}[t]
\centering
    \begin{subfigure}[t]{0.49\linewidth}
    \includegraphics[width=0.99\linewidth]{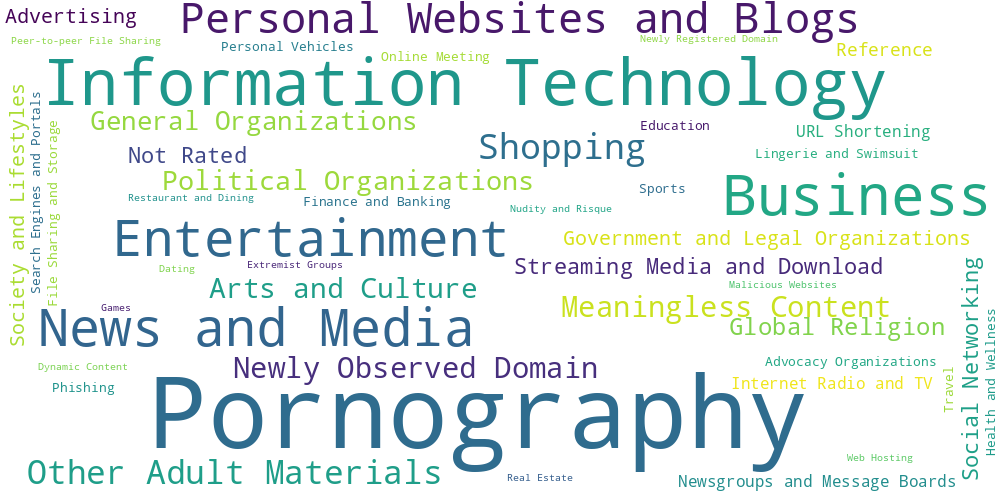}
    \vspace{-6mm}
    \caption{{\small Toxic profiles}}
    \label{fig:top1p_sld_cat_cloud}
    \end{subfigure}
    \hfill
    \begin{subfigure}[t]{0.49\linewidth}
    \includegraphics[width=0.99\linewidth]{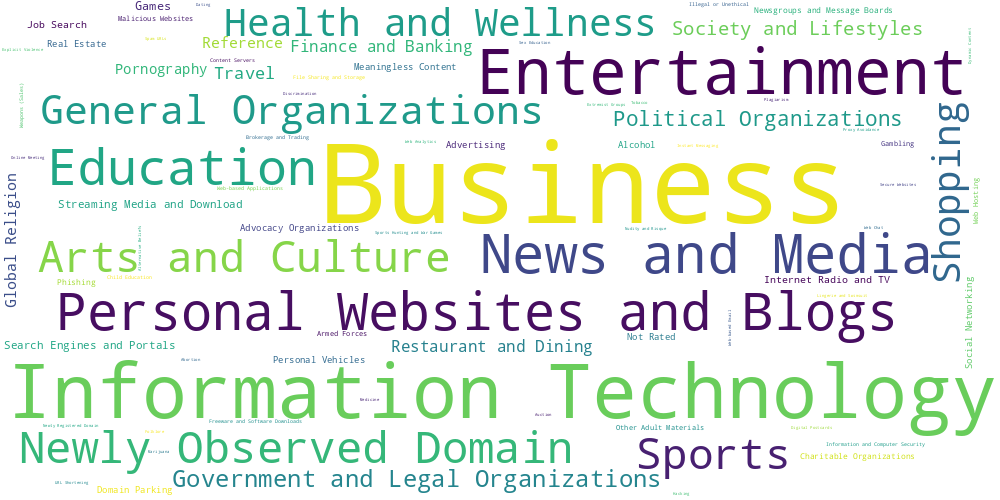}
    \vspace{-6mm}
    \caption{{\small Baseline profiles}}
    \label{fig:baseline_sld_cat_cloud}
    \end{subfigure}
    \vspace{-2mm}
    \caption{\small Second level domains (SLDs) categories shown as word clouds for toxic (\protect\subref{fig:top1p_sld_cat_cloud}) and baseline (\protect\subref{fig:baseline_sld_cat_cloud}) profiles (cf. \S\ref{subsub:url_cats} for details).}
    \vspace{-1mm}
\end{figure}

\subsubsection{What is the nature of categories in the URLs shared from tweets of toxic and baseline profiles?}
\label{subsub:url_cats}

For the URLs that have been shared, the domain can provide an indication of the type of content linked.
For example, {\tt{www.example.com}}'s second level domain (SLD) is {\tt{example.com}}.
Proceeding forwards, ``Domain'' and ``SLD'' are interchangeable. 
We classify the content type of the domain with the \textbf{\emph{FortiGuard}} classification service~\cite{Fortiguard}. FortiGuard uses link crawlers, customer logs, and machine learning to categorize websites~\cite{Triplet20}.
We successfully categorize 98.2\% unique domains for baseline and 95.4\% for toxic profiles with FortiGuard. The total number of found unique categories between the SLDs was 596 for toxic and 1,008 for the baseline. A toxic profile has on average 49 unique SLDs and a baseline profile has 487 unique SLDs in all of the tweets. 
We present in Figs.~\ref{fig:top1p_sld_cat_cloud} and ~\ref{fig:baseline_sld_cat_cloud} a weighted word cloud of the SLD categories, the size of text represents the percentage of SLDs in each group assigned the category label. 
We observe that toxic profiles are linked to domains categorized as Pornography (examples intentionally omitted), and Information Technology (e.g.  \texttt{youtube.com, SelfieSwipes.com, Kailani-Kai.com}). For the baseline profiles, the largest category is business (\texttt{huffingtonpost.com, manchester.ac.uk,gotthevo\newline te.org})
\subsubsection{Do toxic profiles share SLDs about the same subject/topic?}
\label{subsec:sld_similarity}
For this, we now do a direct comparison between the nature of SLDs of toxic and baseline profiles, amongst themselves and between each other. 
Specifically, in Fig.~\ref{fig:jaccard_similarity} we compute and show the CDF of pairwise Jaccard similarity calculated on the sets of SLDs present in each toxic and baseline profile.
The \textbf{\emph{Jaccard Index}} is computed between two sets $A$ and $B$ as $\frac{|A \bigcup B|}{|A \bigcap B|}$, and ranges between 0 (for no common elements between the two sets) to 1 (for a perfect match or overlap). 
We observe that the SLDs in toxic profiles have the greatest overlap. Also, 64\% of pairs of toxic 1\% profiles have no similarity compared to only 18\% of baseline pairs. 44\% of toxic~1\%-baseline pairs have disjoint sets of SLDs, indicating the presence of many SLDs present in toxic 1\% tweets that are absent from baseline tweets. Overall, there is little similarity, with 95\% of baseline-toxic 1\% and baseline-baseline pairs, and 88\% of toxic 1\%-toxic 1\% pairs with Jaccard similarity less than 0.1.

\subsubsection{\bf Takeaways:}
\begin{itemize}[leftmargin=*]
    \item Toxic Profiles use fewer URLs  and generally refer to unique URLs suggesting they refer less to external sources than the rest of the Twitter population.
    \item Of the domains linked by toxic profiles, we observed that the most popular domain categories are pornography, news, and information technology.
    \item Toxic 1\% profiles have a larger proportion of profiles (63.7\%) with no similarity between shared SLDs than baseline profiles (18.7\%). This indicates high uniqueness, either from the independent operation or customized tracking domains.

\end{itemize}

\subsection{Hashtags} 
\label{subsec:hashtag_analysis}
Adding hashtags to tweets is a popular and easy way for users to convey a message to an interested audience, and to have a voice within intended communities.
We compare the tendency of sharing hashtags between the toxic and baseline profiles. 
\subsubsection{Do toxic profiles take the help of hashtags in their tweets?}
\begin{figure}[th]
\centering
    \begin{subfigure}[t]{0.46\linewidth}
    \includegraphics[width=0.95\linewidth]{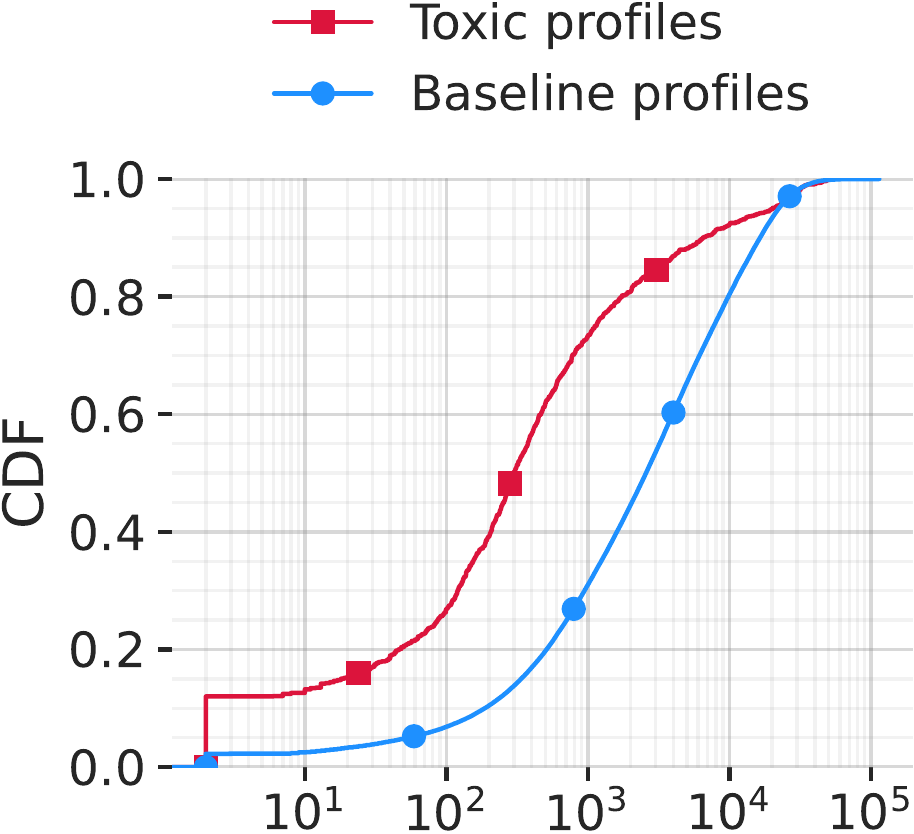}
    \vspace{-0.1mm}
    \caption{{\small \#Hashtags }}
    \label{fig:no_total_hts}
    \end{subfigure}
    \hfill
    \begin{subfigure}[t]{0.46\linewidth}
    \includegraphics[width=0.95\linewidth]{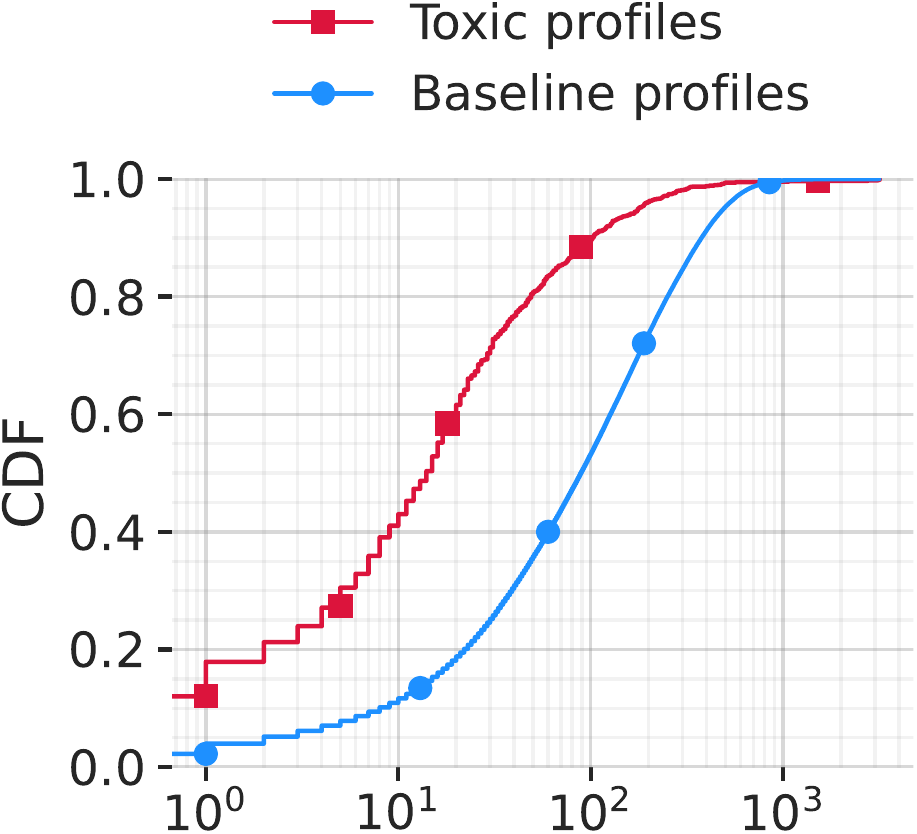}
    \vspace{-0.1mm}
    \caption{{\small\#Unique hashtags}}
    \label{fig:no_unique_hts}
    \end{subfigure}
    \begin{subfigure}[t]{0.46\linewidth}
    \includegraphics[width=0.95\linewidth]{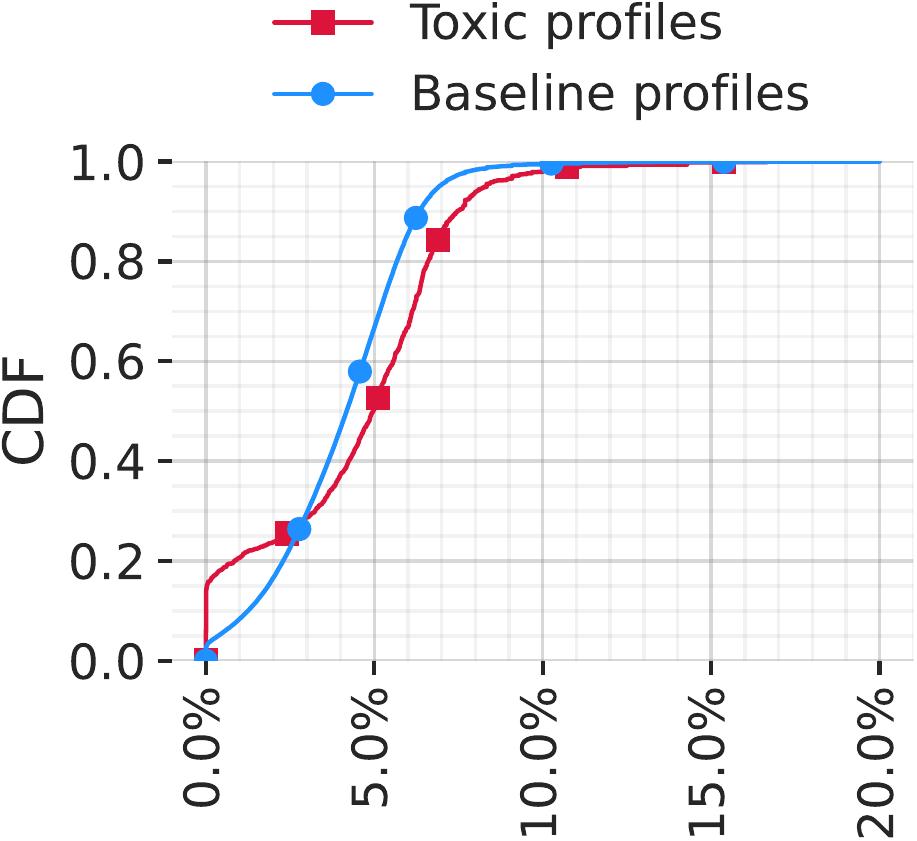}
    \vspace{-0.2mm}
    \caption{{\small \% Unique hashtags}}
    \label{fig:p_unique_hts_combined}
    \end{subfigure}
    \hfill
    \begin{subfigure}[t]{0.46\linewidth}
    \includegraphics[width=0.95\linewidth]{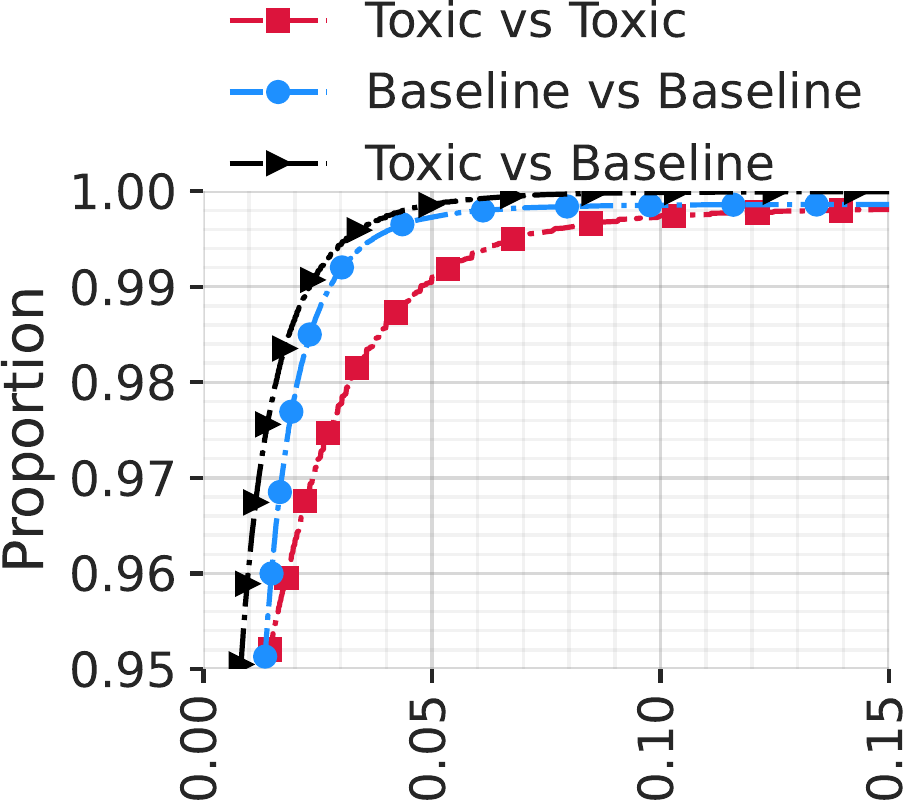}
    \vspace{-0.2mm}
    \caption{{\small Jaccard Similarity (Htags)}}
    \label{fig:jaccard_similarity_hts}
    \end{subfigure}
    \vspace{-2mm}
    \caption{\small Hashtag analysis of toxic and baseline profiles (cf. \S\ref{subsec:hashtag_analysis}).}
    \vspace{-4mm}
\end{figure}

Hashtags place your message within the context of a topic or community.
First, with the total number of hashtags per profile (including repetitions), we discern from Fig.~\ref{fig:no_total_hts} that 50\% of the toxic profiles at least shared 300 hashtags in total but 50\% of baseline profiles shared many more: at least 1150 hashtags. 
Next, on the number of unique hashtags per profile (discounting repetitions), we see from Fig.~\ref{fig:no_unique_hts} that half of the toxic profiles at most shared about 15 unique hashtags, and half of the general profiles shared at most 100 unique hashtags. 
As such it is evident that the toxic users are using hashtags less than the baseline.
Next, Fig.~\ref{fig:no_unique_hts} shows us that the toxic profiles also share fewer unique hashtags than the baseline. Fig.~\ref{fig:p_unique_hts_combined} tells us that half of the toxic profiles at most shared 50\% unique hashtags as compared to 60\% of baseline. Jaccard similarity Fig.~\ref{fig:jaccard_similarity_hts}
of the hashtags shows us the strongest similarity amongst the hashtags of toxic profiles again, pointing to their narrow focus as observed through a small number of unique categories of SLDs in \S\ref{subsec:sld_similarity}.

\begin{figure}[t]
\centering
    \begin{subfigure}[t]{0.49\linewidth}
    \includegraphics[width=0.99\linewidth]{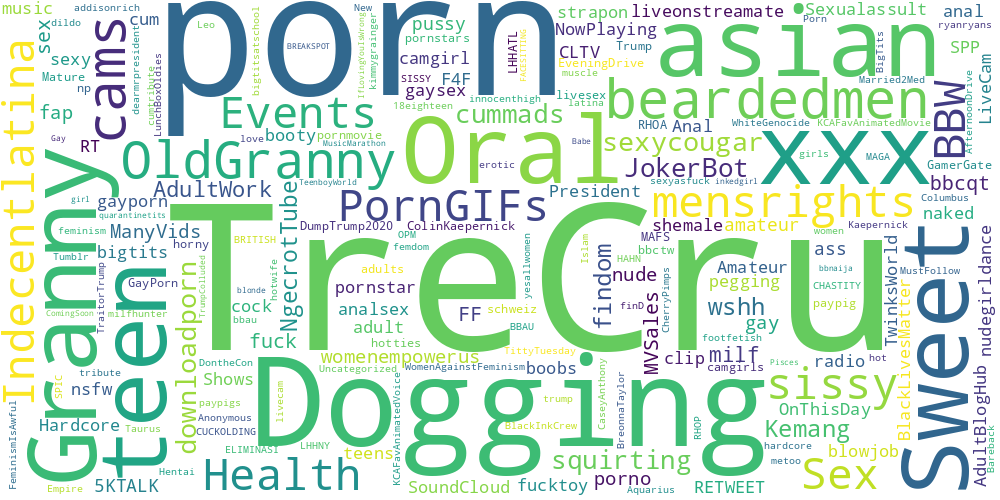}
    \vspace{-4mm}
    \caption{{\small Toxic profiles}}
    \label{fig:hts_cloud_top1p}
    \end{subfigure}
    \hfill
    \begin{subfigure}[t]{0.49\linewidth}
    \includegraphics[width=0.99\linewidth]{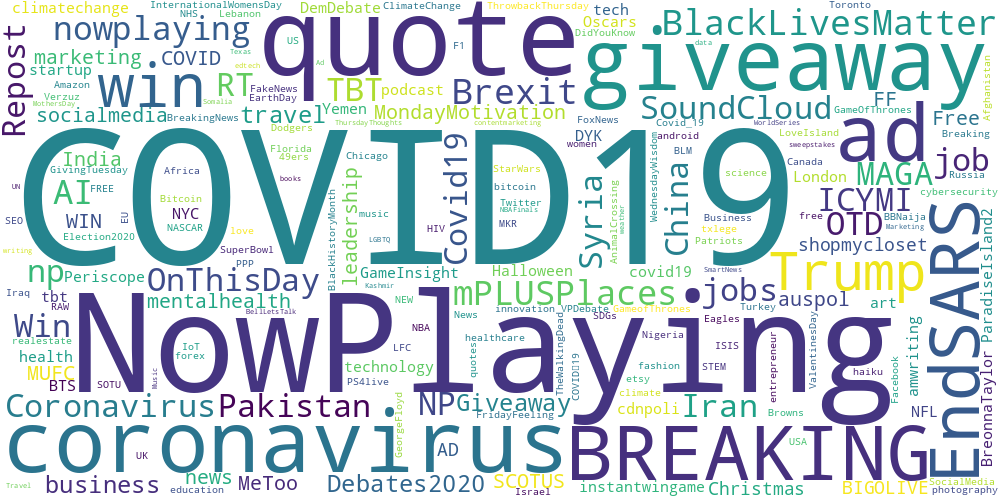}
    \vspace{-4mm}
    \caption{{\small Baseline profiles}}
    \label{fig:hts_cloud_baseline}
    \end{subfigure}
    \vspace{-2mm}
    \caption{\small Hashtag word clouds of profile groups (cf. \S\ref{subsub:hashtag_cat}).}
    \vspace{-4mm}
\end{figure}
\subsubsection{Nature of hashtags shared by the top 1\% toxic and baseline profiles}
\label{subsub:hashtag_cat}

We now provide examples of highly occurring hashtags within the dataset. We present in 
Fig.~\ref{fig:hts_cloud_top1p} and ~\ref{fig:hts_cloud_baseline} the weighted word clouds of all hashtags collected from tweets of toxic and baseline profiles. The largest hashtag shared by toxic profiles is 'TreCru` which is an online video game known as Treasure Cruise. The remainder of the hashtags by toxic profiles are of a very explicit nature. On the other hand, the general Twitter profiles share hashtags about diverse topics including Covid, news, and politics such as \#BlackLivesMatter, \#Trump, \#Brexit, \#EndSARS which is a protest against police brutality in Nigeria. There are also everyday benign hashtags about music \#nowplaying, \#SoundCloud and shopping \#Giveaway, \#Job. 
On average toxic profiles share 59 unique hashtags per profile vs 275 by a baseline profile.
 
\subsubsection{Do toxic profiles share the same or similar hashtags in their tweets?}
\label{subsec:hash_similarity}

By leveraging the same Jaccard similarity metric defined in \S\ref{subsec:sld_similarity}, we inspect the overlapping hashtags used within and between the toxic and baseline profiles. While not visible in Fig.~\ref{fig:jaccard_similarity_hts}, it is noted that there is zero hashtag similarity in 90.2\% of toxic--toxic profiles, 85.5\% between toxic-baseline profiles, and 62.9\% of baseline--baseline profiles. What is visible in Fig.~\ref{fig:jaccard_similarity_hts} illustrates that a small proportion of toxic-toxic profiles have a much higher overlap of hashtags and thus an aligned area of discussion.
 
\subsubsection{\bf Takeaway:}
 Toxic profiles share fewer total and unique hashtags than baseline profiles. A larger majority of toxic profiles do not have overlapping hashtags with other toxic profiles (90.2\%), compared to the baseline (62.9\%). In the toxic profiles that do share hashtags with other toxic profiles, they are more aligned than the most overlapping baseline-baseline profiles.

\begin{figure*}[t]
    \centering
    \begin{minipage}{.32\linewidth}
    \includegraphics[width=0.95\linewidth]{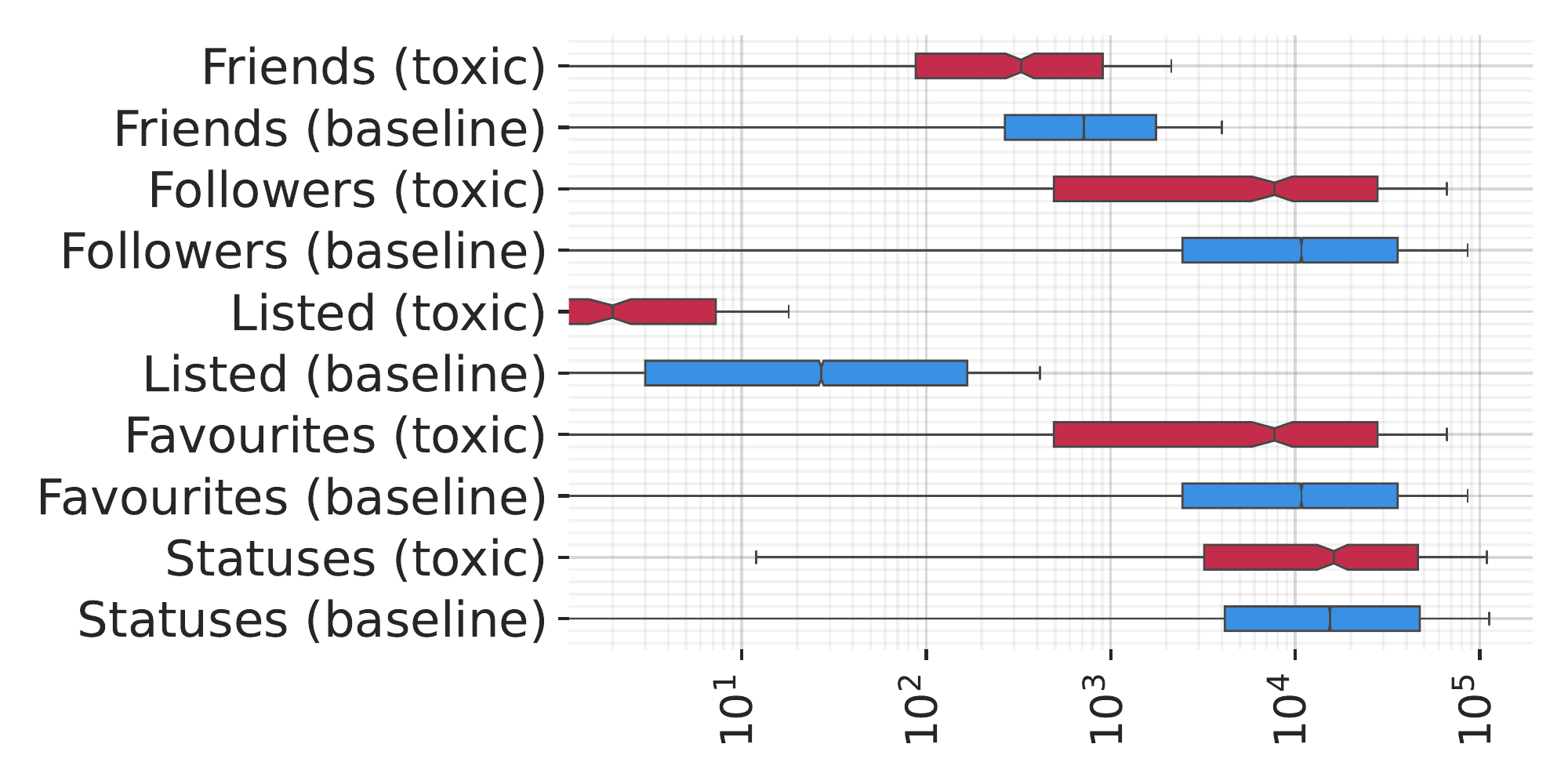}
    \vspace{-2mm}
    \caption{\small Toxic and baseline profiles' features (e.g. \#friends, \#followers, etc.)
    (cf. \S\ref{sec:profile_data_analysis})}
    \label{fig:profiles_data}
    \vspace{-1mm}
    \end{minipage}
    \hfill
\begin{minipage}{.33\linewidth}
    \centering
    \includegraphics[width=0.95\linewidth]{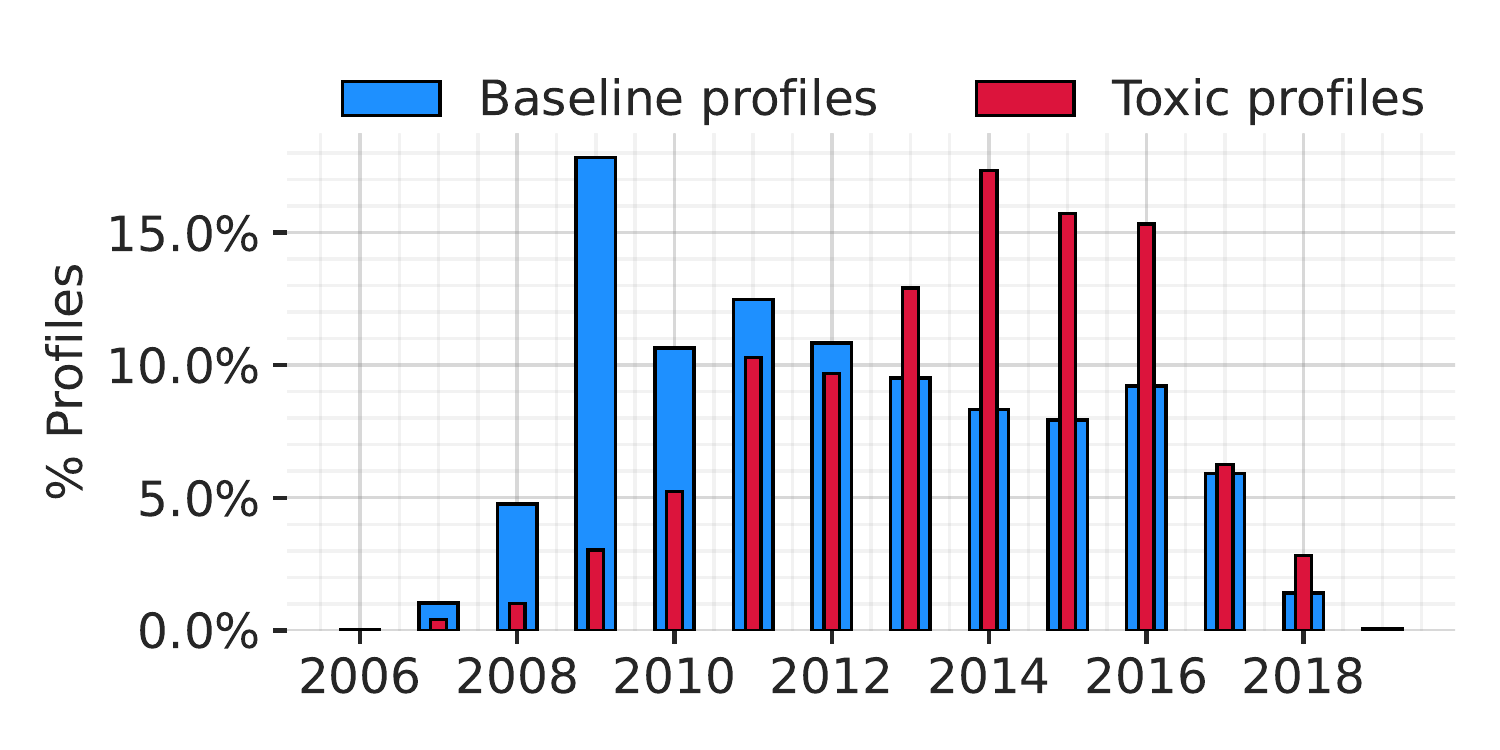}
    \vspace{-1mm}
    \caption{\small Creation dates of toxic and baseline profiles (cf. \S\ref{sec:profile_data_analysis}).}
    \label{fig:profile_age}
    \vspace{-1mm}
\end{minipage}
\hfill
\begin{minipage}{.33\linewidth}
    \centering
    \includegraphics[width=0.98\linewidth]{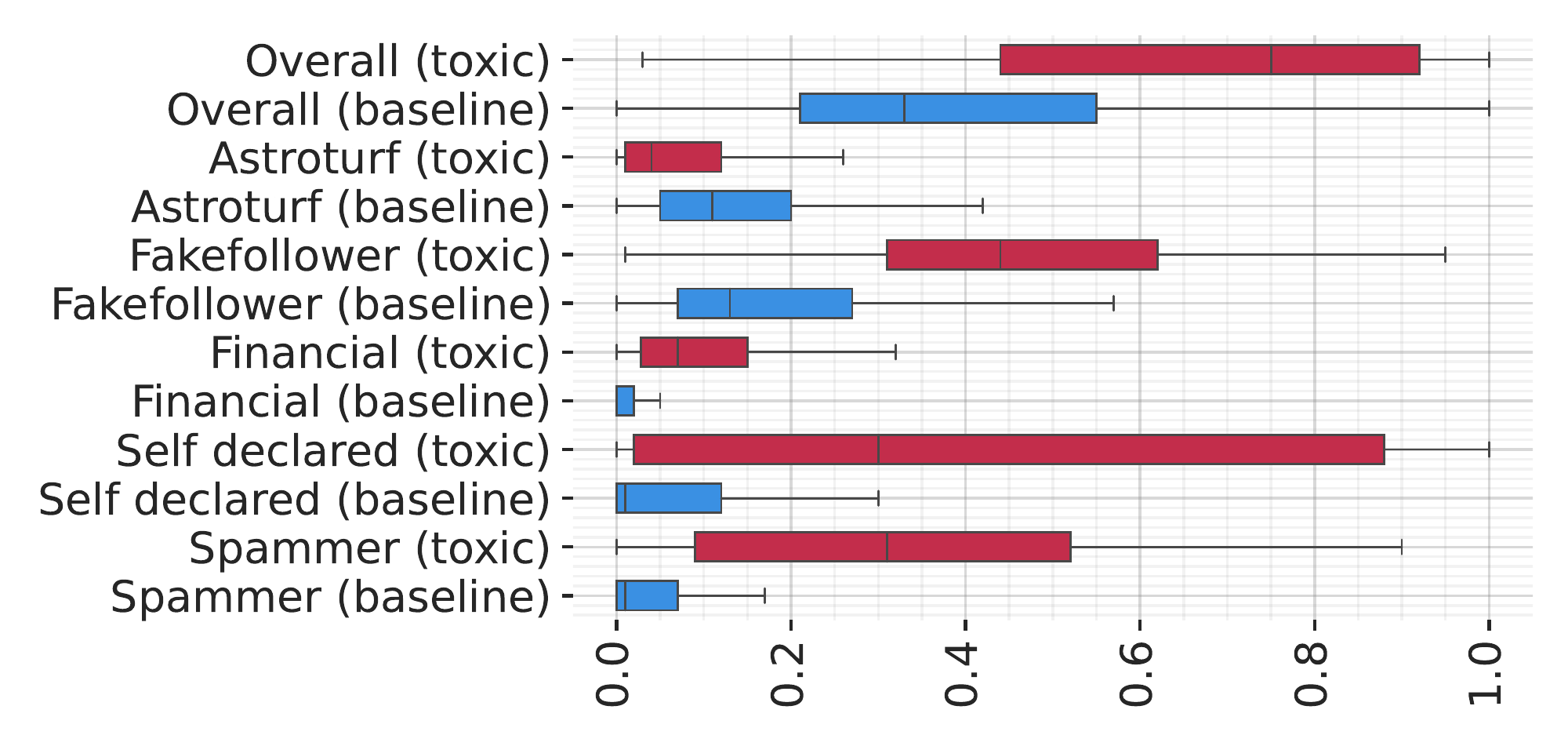}
    \vspace{-1mm}
    \caption{\small Botometer automation scores of toxic and baseline profiles (cf. \S\ref{sec:automation_analysis}).}
    \label{fig:botometer_scores}
    \vspace{-1mm}
    \end{minipage}
\end{figure*}

\section{Profile Level Analysis}
\label{sec:profile_analysis}
In this section, we observe profile-level characteristics that emerge from our toxic and baseline profiles. We shall start by analyzing details directly registered with Twitter (\S\ref{sec:profile_data_analysis}), followed by an analysis of automation as provided by the Botometer (\S\ref{sec:automation_analysis}).
\subsection{Account Metadata}
\label{sec:profile_data_analysis}
`Metadata is a ``data dictionary'' attached to every Twitter profile providing additional insight about a profile. It is a dictionary of 17 fields including name, location, account creation date, counts of friends, followers, statuses, and favorites. It also contains information if an account that is protected and/or verified.

\subsubsection{What does a toxic profile's Twitter account metadata say about them?}
We first inspect the proportion of profiles that are still present on Twitter. It is seen from Tab.~\ref{tab:twitter_profile} that all toxic profiles are still present on Twitter to this day, whereas 3.6\% of baseline profiles no longer exist. We unfortunately cannot further determine if these accounts were deleted or banned.
A majority of toxic profiles are verified; \emph{A profile with a blue badge to show that an account is Twitter verified}. Twitter allows automation~\cite{twitter_automation} and verifies the bot account~\cite{bot_verification}, it, however, does not allow misconduct. Also, 95\% of the Twitter profiles are protected; \emph{A profile that does not appear in third-party search engines, i.e, Google,~\cite{twitterAPI}}. No toxic profiles are banned in any specific country, whereas 0.002\% of baseline profiles are in multiple countries like Russia, Austria, and Belgium to name a few.
Additional numeric data is provided in the Twitter profile object, and the distribution of the values is represented in Fig.~\ref{fig:profiles_data}. 

\textbf{``Friends and Followers''} Friends of a Twitter profile are other profiles followed by said profile. While followers are other profiles that follow the said profile. We observe a toxic profile has on average 500 friends, and 9,500 followers, whereas the average baseline profiles have 800 friends and 10K followers in total. 

\textbf{``Listed''} gives the number of public lists that this profile is included within, these lists are used to collect similar accounts to strategize the timelines.
As few as a couple of toxic profiles exist as part of any list, whereas a median baseline profile on average is included within 50 lists. 

\textbf{``Favourites''} is the number of Tweets liked by a profile. We note that an average profile in both groups has liked an equally significant number of tweets (10K). 

\textbf{``Statuses''} are the number of tweets (including retweets) created by the profile in totality, beyond the 3200 restrictions imposed during the crawling of a profile's timeline. It is interesting to note that the number of tweets posted by a median toxic and baseline profile is approximately the same if all the historical tweets are taken into account.

\textbf{``Location''}
Another dimension is the reported location of the profile. This is a text string typically populated with a description of the profiles home town, state, and/or country. We leverage ``Geopy''~\cite{Geopy:Documentation}, a python library to resolve these strings into country names.
The countries with the highest occurrence are presented in Tab.~\ref{tab:twitter_profile_location}, only the top 3 countries are shown due to the quantity of found countries. 

We can observe that the majority of toxic and baseline profiles are based in the US, UK, and Canada, but there exists a long tail of other countries with which profiles are associated. Interestingly, the toxic profiles are more strongly concentrated in the US (61.36\%) compared to the proportion of baseline profiles in the US (29.42\%). This finding is likely to differ when a different language is considered. We also note that only 0.04\% of toxic and 14\% of baseline profiles enabled ``Geolocation'' in their profile.

\textbf{``Creation date''}
Finally, we inspect the creation date of the profiles in Fig.~\ref{fig:profile_age}. We observe that toxic profiles skew younger than the baseline profiles. A possible explanation is an increase in the creation of toxic profiles around 2014-2016, as observed by~\cite{kollanyi2016bots}. We acknowledge that the forced and voluntary deletion of toxic profiles may also bias these numbers. Interestingly, there was a notable decline in the growth of profiles after 2016, which coincides with a plateau of active users on Twitter~\cite{monthlyactiveusers}.

\begin{table}[t]
\resizebox{\columnwidth}{!}{
\begin{tabular}{l|cccc}
\toprule
&\bf Active profiles & \bf Protected & \bf Verified & \bf Withheld in countries \\ \midrule
\bf Toxic Profiles&  100\%& 92.74\% & 96.5\%& None \\ \hline
\bf Baseline Profiles&  96.4\% &95.74\% &82.6\% & 0.002\%\\
 \bottomrule
\end{tabular}
}
\caption{\small Twitter profiles data (cf. \S\ref{sec:profile_data_analysis}).}
\label{tab:twitter_profile}
\vspace{-4mm}
\end{table}

\begin{table}[t]
\resizebox{\columnwidth}{!}{
\begin{tabular}{l|l}
\toprule
& \bf Top 3 locations found in profiles  \\ \midrule
\bf Toxic Profiles& US(61.36\%), Canada(9.09\%), UK(9.09\%), 9 others(20.04\%)  \\ \hline
\bf Baseline Profiles& US(49.42\%), UK(22.61\%), Canada(7.0\%), 33 others(20.09\%)  \\
 \bottomrule
\end{tabular}
}
\caption{\small Twitter profiles location (cf. \S\ref{sec:profile_data_analysis}).}
\label{tab:twitter_profile_location}
\end{table}
\subsubsection{\bf Takeaways:}
\begin{itemize}[leftmargin=*]
    \item Toxic profiles, in general, have fewer friends, and followers and are not part of public lists in other accounts. Of the toxic profiles that have a location, 61\% of them are based in the US. 
    \item We observe an increase in the creation of toxic profiles in the US election years between 2014 and 2016, matching previous work. 
\end{itemize}

\subsection{Automation}
\label{sec:automation_analysis}
\subsubsection{Are toxic profiles automated bots?} Automated accounts or ``bots'' have been observed on Twitter~\cite{TwitterBotAccounts}, however, Twitter permits automated accounts when they behave well according to Twitter's policy~\cite{Twitterautomation}. 
Thus, in addition to investigating the percentage of bots in our toxic and baseline groups, we also scrutinize the percentage of Twitter policy breaching ``bad bots'', e.g. Spammer, Fake Follower bots from \emph{Botometer API v4}~\cite{Sayyadiharikandeh_2020}. 
The Botometer API provides scores from five classifiers that estimate a profile's similarity to different kinds of bot behavior, including Fake Follower bots, Financial bots, self-declared bots, spammer bots, and astroturf accounts. 
Botometer API leverages features of a profile including the number of friends, social network structure, temporal activity (e.g. tweeting, likes, retweets), tweet language, and sentiment.
Botometer provides scores in the [0,1] range, using either English or Universal (language-independent) features (we report overall universal feature scores). 
Botometer API defines each as:
\begin{itemize}[leftmargin=*]
    \item {\bf Bot score}: An overall probability of profile being a bot
    \item {\bf Astroturf}: A profile being one of the manually labeled political bots. These accounts systematically delete content over time.
    \item {\bf Fake Follower}: An account being a bot purchased to increase follower counts.
    \item  {\bf Financial}: A profile used to post cashtags. Cashtags are stock symbols used with the ``\$'' symbol. Cashtags bots promote low-value stocks by exploiting the popularity of high-value ones.
    \item  {\bf Self Declared}: A profile that is a bot registered with \texttt{botwiki.org}.
    \item {\bf Spammer}: A profile labeled as spam bots from several datasets.
\end{itemize}
The scores for every profile are presented in Fig.~\ref{fig:botometer_scores}.
Each boxplot details the mean and standard deviation of all scores for both toxic and baseline profiles. The scores range from [0-1], with 0 being the most human-like and 1 as the most bot-like. 

We observe that toxic profiles generally have higher overall Botometer scores with a median of 0.7, however, there still exists toxic profiles that are human-like with scores in the 1st standard deviation range <0.45 overall score. Baseline profiles skew more human-like in comparison. 
Astroturf (participating in politics \cite{ratkiewicz2011detecting}) scores are fairly low for both sets, albeit baseline profiles skewing slightly higher, this may explain that despite the most toxic these profiles are not automatically removed by Twitter, there are still long-lived spam and profane accounts on Twitter, also reported by~\cite{6488315}.
Very few of the baseline profiles are likely to be fake followers, with a median probability of 0.14. On the other hand, just under half of the toxic profiles have a probability above 0.5 and are likely to be purchased, followers. This indicates the presence of maliciously toxic actors amplifying their toxic message through these profiles. 
Neither set of profiles are likely to engage in financial market updates, though there are notably more among toxic profiles. 
We observe that toxic profiles are widely spread on the spectrum of ``self-declared'' in stark contrast to baseline profiles. Spamming is not a trait of baseline profiles whereas toxic profiles have notably higher scores, and there are toxic profiles with spammer scores as high as 0.85. 
\subsubsection{\bf Takeaways:}
\begin{itemize}[leftmargin=*]
    \item We confirm the findings reported in ~\cite{doi:10.1080/10584609.2018.1526238} that the distribution of toxic profiles is less likely to be associated with politics, despite their toxic nature.  %
    \item Toxic and baseline profiles are unlikely political or financial bots.  %
    Our study validates prior work~\cite{6488315, Twitterverifiedbots} that the toxic profiles have a high likelihood to be spam bots and have behavior consistent with self-declared bots. 86.5\% of toxic profiles are verified (\S\ref{sec:profile_data_analysis})--as also reported in~\cite{Twitterverifiedbots}. 
    \item Many validated toxic profiles are verified which makes their content more viral as also found by~\cite{mathew2018analyzing}.
\end{itemize}

\section{Ethical Considerations}
Our research is non-commercial, and in line with Twitter’s Terms and Conditions for research purposes, our data will not be shared with any third party for commercial purposes.
We used the standard Twitter API to collect tweets from public user profiles. In all of our experiments, any result produced and shown cannot be used to re-identify, or track said users, as no user profiles are specifically named. During our experiments, we follow ethical guidelines outlined in~\cite{rivers2014ethical}.
Given our experimentation on human-produced data, we obtained IRB approval\footnote{Reference no: 520211000835379} from our institution.

\section{Discussion}
Understanding the prevalence of hateful information on social media platforms is the primary driver behind this investigation of the most toxic profiles.
In order to characterize levels of consistency of such behavior, to be able to make early predictions of the spread of such content, and in essence to prevent the proliferation of the most hateful content providers, it would be beneficial to study the population of profiles who produce the most toxic content.  

Being able to adequately understand the behavior of the most toxic Twitter profiles is valuable in and of itself. It provides more well-informed choices about how and what to research in subsequent research investigations.
It makes it possible for toxicity-reduction strategies to be designed more intelligently in many ways.
  
We concede that we had to work within certain constraints, such as the Perspective API limiting our work to only English tweets.
Also, the Twitter API only allowed us to scrape the most recent 3200 tweets and not the entire timeline per profile. We will also like to acknowledge that because our seed data were balanced, with equal amounts of toxic and non-toxic profiles, the 1\% is not a complete portrayal of the entire Twitter-sphere.  

In the future, we plan to use our findings of highly toxic Twitter profiles to identify toxic Twitter profiles that are responsible for the highest toxicity in important Twitter discussions about politics, sports, and religion, among others. 
\label{sec:Discussion}
\section{Conclusion}
\label{sec:discussion_and_conclusion}

In the past, much research has been devoted to locating toxicity spreader accounts and bots based on a few tweets; however, our work examines the timeline of Twitter accounts and takes into account the consecutive tweets posted by Twitter accounts, as well as investigates the consistent toxicity exhibited by certain profiles.

We present a profile-centric approach to survey toxicity on Twitter and characterize the most toxic profiles. Our methodology is distinct from prior works that focus on particular events or hashtags and phrases over short time windows. We focus instead on the entire profile timeline, obtaining longitudinal data that reveals 
the bigger picture of a profile's toxic behavior. We annotate entire timelines with the Google Perspective API. Based on the toxicity of profile tweets, we isolate the 1\% most toxic profiles in our dataset and contrast their behavior with the remainder. 
This approach provides extra context to a profile's toxic behavior, providing new insights into toxicity on Twitter. 

We find that the most toxic 1\% of profiles are likely to be fake followers, indicating a level of coordinated and targeted toxic activity. They are likely to post inflammatory and profane content. Their tweets are typically textually eloquent and tend to repeat their posted content less. They are less likely to leverage auxiliary content such as URLs and hashtags in their posts.

Inspecting toxic profiles on their longitudinal data provides additional contextual insights that are otherwise missing when scrutinizing a profile on a single post. However, our approach still has limitations as obtaining this data is a challenging task. Specifically, Twitter limits the availability of the timeline to a profile's 3,200 most recent tweets. Certainly, with the full timeline, further insights could be obtained.

Findings regarding characteristics of the most toxic profiles such as
inflammatory and insulting behavior, repetitive and explicit hashtags, bursty tweeting patterns, and short and well-written tweets with supporting URLs to websites and blogs in their tweets can be used to identify the most toxic Twitter accounts in a specific scenario, such as profiles discussing politics or following a specific motivational movement. Furthermore, identifying and deleting such accounts will aid in the removal of toxicity from important Twitter discussions. Our approach is not limited to Twitter and can be applied to any social media platform discussion.

In the future, we plan to further study the details of the topics discussed by toxic profiles and investigate and characterize the coordinated toxic activity as evidence for toxic influence operations present in the data.

\begin{acks}
This work was partially supported by the Macquarie University Cybersecurity Hub (MQCHUB) and the EU H2020 Research and Innovation program. Hina Qayyum was supported by Macquarie University Domestic High Degree Research Scholarship Program. Nicolas Kourtellis was  partially supported during this project by the EU H2020 Research and Innovation program under grant agreements No 830927 (Concordia), 871793 (Accordion), and 101021808 (Spatial). Any opinions, findings, conclusions, or recommendations expressed in this material are those of the authors or originators and do not necessarily reflect the views of the MQCHUB or of the EU H2020 Research and Innovation program.
\end{acks}

\bibliographystyle{ACM-Reference-Format}
\bibliography{reference}
\appendix
\end{document}